\documentclass[%
 reprint,
 amsmath,amssymb,
 aps,
prstab,
]{revtex4-1}

\usepackage{graphicx}% Include figure files
\usepackage{dcolumn}% Align table columns on decimal point
\usepackage{bm}% bold math
\usepackage{stmaryrd}
\usepackage{hyperref}% add hypertext capabilities
\begin{document}

\preprint{APS/123-QED}

\title{Low power commissioning of an innovative laser beam circulator\\ for inverse Compton scattering $\gamma$-ray source}% Force line breaks with \\

\author{Cheikh Fall Ndiaye}
\altaffiliation[Also at ]{ALSYOM, Tarbes, France}
\affiliation{LAL, Univ. Paris-Sud, CNRS/IN2P3, Universit\'e Paris-Saclay, Orsay, France
}%

\author{Nicolas Beaug\'erard}
\affiliation{%
SEIV, M\'erignac, France
}%

 \author{Bastien Lacrampe}
\affiliation{%
ALSYOM, Tarbes, France
 }%

\author{Herv\'e Rocipon}
\affiliation{%
ALSYOM, Tarbes, France
 }%
 
\author{Kevin Cassou}
 \email{cassou@lal.in2p3.fr}
 \homepage{http://www.lal.in2p3.fr}
\affiliation{%
LAL, Univ. Paris-Sud, CNRS/IN2P3, Universit\'e Paris-Saclay, Orsay, France
}%

\author{Patrick Cornebise}
\affiliation{%
LAL, Univ. Paris-Sud, CNRS/IN2P3, Universit\'e Paris-Saclay, Orsay, France
 }%

\author{Kevin Dupraz}
\affiliation{%
LAL, Univ. Paris-Sud, CNRS/IN2P3, Universit\'e Paris-Saclay, Orsay, France
 }%

 \author{Denis Douillet}
\affiliation{%
LAL, Univ. Paris-Sud, CNRS/IN2P3, Universit\'e Paris-Saclay, Orsay, France
 }%
 
 \author{Titouan Le Barillec}
\affiliation{%
LAL, Univ. Paris-Sud, CNRS/IN2P3, Universit\'e Paris-Saclay, Orsay, France
 }%
 
 \author{Christopher Magueur}
\affiliation{%
LAL, Univ. Paris-Sud, CNRS/IN2P3, Universit\'e Paris-Saclay, Orsay, France
 }%
 
\author{Aur\'elien Martens}
\affiliation{%
LAL, Univ. Paris-Sud, CNRS/IN2P3, Universit\'e Paris-Saclay, Orsay, France
 }%
 
   \author{Daniele Nutarelli}
\affiliation{%
 LAL, Univ. Paris-Sud, CNRS/IN2P3, Universit\'e Paris-Saclay, Orsay, France
 }%

 \author{Yann Peinaud}
\affiliation{%
LAL, Univ. Paris-Sud, CNRS/IN2P3, Universit\'e Paris-Saclay, Orsay, France
 }%
 
 \author{Alice Thi\'ebault}
\affiliation{%
LAL, Univ. Paris-Sud, CNRS/IN2P3, Universit\'e Paris-Saclay, Orsay, France
 }%

 \author{Themis Williams}
\affiliation{%
LAL, Univ. Paris-Sud, CNRS/IN2P3, Universit\'e Paris-Saclay, Orsay, France
 }%

  \author{Fabian Zomer}
\affiliation{%
 LAL, Univ. Paris-Sud, CNRS/IN2P3, Universit\'e Paris-Saclay, Orsay, France
 }%

       \author{David Alesini}
\affiliation{%
LNF-INFN, Via E. Fermi 40, 00044, Frascati, Roma, Italy
 }%
 
        \author{Fabio Cardelli }
\affiliation{%
LNF-INFN, Via E. Fermi 40, 00044, Frascati, Roma, Italy
 }%
 
       \author{Antonio Falone}
\affiliation{%
LNF-INFN, Via E. Fermi 40, 00044, Frascati, Roma, Italy
 }%
 
        \author{Giovanni  Franzini}
\affiliation{%
LNF-INFN, Via E. Fermi 40, 00044, Frascati, Roma, Italy
 }%
 
    \author{Alessandro Gallo}
\affiliation{%
LNF-INFN, Via E. Fermi 40, 00044, Frascati, Roma, Italy
 }%
 
  \author{Luca Piersanti}
\affiliation{%
LNF-INFN, Via E. Fermi 40, 00044, Frascati, Roma, Italy
 }%
 
      \author{Valerio Petinacci}
\affiliation{%
LNF-INFN, Via E. Fermi 40, 00044, Frascati, Roma, Italy
 }%
 
    \author{Stefano Pioli}
\affiliation{%
LNF-INFN, Via E. Fermi 40, 00044, Frascati, Roma, Italy
 }%
 
   \author{Alessandro Variola}
\affiliation{%
LNF-INFN, Via E. Fermi 40, 00044, Frascati, Roma, Italy
 }%
 
    \author{Andrea Mostacci}
\affiliation{%
Universit\`a di Roma "La Sapienza", Dipartimento Energetica, Via A. Scarpa 14, 00161, Roma, Italy
 }%

  \author{L. Serafini}
\affiliation{%
INFN-Mi, via Ceriola 16, 20133, Milano, Italy
 }%

\collaboration{EuroGammaS Consortium}%\noaffiliation

\date{\today}% It is always \today, today,

\begin{abstract}
We report on the optical commissioning of the high power laser beam circulator (LBC) for the high brightness Compton $\gamma$-ray source Extreme Light Infrastructure for Nuclear Physics. Tests aiming at demonstrating the optical performances of the LBC have been realized with a low-power pulsed laser-beam system and without electron beam. We show that, with the developed alignment and synchronization methods coming from the LBC design study presented in the Dupraz \textit{et al.} paper \cite{Dupraz:2014aa}, the LBC enhances the laser-beam power available at the interaction point (IP) by a factor in excess of 25. This corresponds to a potential of bringing the average laser-beam power in excess of $1\,$kW when the LBC is injected with the interaction point laser-beam pulse energy of $400\,$mJ at $100\,$Hz. 

\end{abstract}

\pacs{Valid PACS appear here}% PACS, the Physics and Astronomy
\maketitle

%\showthe\columnwidth

\tableofcontents

\section{\label{sec:introduction} Introduction}

Nowadays,  several facilities around the world produce intense  $\gamma$-ray with various excellent properties such as high flux, low bandwidth, energy tunablity, and high polarization, mainly for applications in nuclear physics and non-destructive analysis \cite{Aoki:2004aa,Weller:2009aa}.  Such  $\gamma$-ray sources are generated by the inverse Compton scattering, process whereby a laser photon beam scatters off a relativistic electron resulting an energy upshifted scattered photon beam. The phase space density of the electron and laser beams at the interaction are key parameters \cite{Sprangle:1992aa,Tomassini:2005aa,Tomassini:2008aa,Petrillo:2012aa} that need particular care in order to achieve high flux along with very small bandwidth of the $\gamma$-ray beam, typically requested by nuclear physics experiments and the related societal applications \cite{VanNoorden:2013aa, Luo:2016aa, Kneissl:1996aa}. An advanced $\gamma$-ray source is being built in Romania for the Nuclear Physics Pillar (ELIN-NP) in the framework of the European Laser Infrastructure \cite{Adriani:2014aa,Giribono:2017aa} by the EuroGammaS consortium \cite{EuroGammaS}. 

The $\gamma$-rays will be generated by a high quality electron bunch train colliding with a high-power laser-pulse that recirculates 32 times at the interaction point. The Gamma Beam Source (GBS) is expected to produce photons with tunable energy between $0.2$ and $19.5\,$MeV with a narrow bandwidth below $ 0.5\%$ and a high spectral density  of $0.8-4 \times 10^4\,$ ph/s eV \cite{Bacci:2013aa}. The GBS will consist of a low-energy  $\gamma$-ray beam line and a high-energy $\gamma$-ray beam line covering the $0.2-3.5\,$MeV and $3.5-19.5\,$MeV ranges photon energies, respectively.  

Beyond the state of the art level requirement on the electron beam accelerator and the $\gamma$ beamline, the interaction point is one of the bottleneck of the GBS. High flux and high brilliance inverse Compton scattering are based electron storing ring coupled to high finesse resonant Fabry-P\'erot optical cavity.  Stored laser power up to hundred's of kiloWatt have been demonstrated in enhanced resonant optical cavity \cite{Carstens:2014aa}, but it remains insufficient to reach the required $\gamma$-ray flux keeping, the low bandwidth and  continuous energy tunability.  Recent work have shown an interesting option, combining the advantage of the laser power amplification of an enhanced resonant optical cavity seed in burst-mode and a room-temperature electron linac delivering a microsecond length bunch train \cite{favier:2018aa}. But, the minimal time spacing of $15\,$ns between each $\gamma$-ray pulse required for the GBS ruled out this optical configuration for the interaction point. Multi-pass non-resonant cavity studies have been carried out for inverse Compton scattering source \cite{Rollason:2004aa, Jovanovic:2007aa,Yamane:2010aa,Chaleil:2016aa} and have shown to be adapted to low number of bunches and tens of nanoseconds circulation period. The multi pass non-resonant cavity or laser beam circulator designed for the GBS has got the particularity to achieve constant interaction parameter (angle, waist, polarization) at a fixed interaction point all along the passes.  

% removed reference Vaccarezza:2014aa ... problem to be  checked

In this article, the optical commissioning with a low power laser system, and without the linac, of the first interaction point module corresponding to the low-energy $\gamma$-ray beam line of the GBS is described. A comprehensive review of the first construction, implementation and commissioning of the new system constituted of the LBC is given. The full GBS is quickly described offering a complete view of the machine in the section \ref{sec:GBS}.  In section \ref{sec:IPmodule}, the IP module elements, their individual testing and characterization are described. In section \ref{sec:TestConditions}, details of the test conditions are given. The presentation of the LBC pre-alignment procedure and results follows in section \ref{sec:LBCprealignment}. The obtained performances are described in Section \ref{sec:LBCresults}. Finally a summary is given in section \ref{sec:Summary} along with insights on the expected performances for the integration of the LBC in the GBS in Romania.

\section{\label{sec:GBS} Gamma Beam Source machine description}

The GBS linear electron accelerator layout is based on a S-band photo-injector at $\nu_{rf}=2856\,$MHz and C-band rf linac, at $5712\,$MHz, similar to that of SPARC$\_$LAB at LNF-INFN, Frascati, Italy \cite{Ferrario:2010aa}. It has the capacity to deliver a high phase-space density electron-beam with energies in the $80-740\,$MeV range. The repetition rate of the pulsed rf is $100\,$Hz. In order to reach the required photon flux, the machine is foreseen to work in a multi-bunch mode.  The 32 electron bunches spaced in time by $T=46/\nu_{rf} =16.1\,$ns are produced in each rf pulse. The effective repetition rate of the collisions is thus $3.2\,$kHz. The main electron beam parameters of the ELI-NP-GBS, along with the description of the complete rf system have been reported in Refs.~\cite{Piersanti:2016aa,Giribono:2017aa}.
The preservation of the electron beam quality is ensured by an hybrid scheme consisting in a S-band high brightness photoinjector followed by a C-band rf linac. The reason lies in having a bunch long enough, $\sigma_z \approx 1\,$mm, in the rf gun to limit the emittance degradation due to the space charge contribution, but taking advantage of the higher accelerating gradients provided by the C-band accelerating sections in the rest of the linac to keep it compact. To avoid the energy spread dilution due to rf curvature degradation effects, a bunch as short as $\sigma_z\approx 280\,\mu$m, is injected in the C-band linac. This is done by means of a velocity bunching scheme implemented in the S-band injector. The S-band injector is composed of a $1.6$ cell S-band rf gun equipped with a copper photocathode and an emittance compensation solenoid, followed by two travelling wave SLAC type S-band sections. The rf gun accelerating field is $E_{acc} \sim 120\,$MV/m, while the two S-band structures can operate at a maximum of $23.5\,$ MV/m giving a maximum electron beam energy at the injector exit of $146\,$MeV \cite{Alesini:2015aa}.
The downstream C-band rf linac operates at $5.712\,$GHz, with the accelerating structures designed and developed at LNF \cite{Alesini:2017aa}, where the accelerating gradient can be set up to a maximum of $33\,$MV/m, providing enough margin for the off crest minimization of the energy spread in the $75-740\,$MeV energy range. The C-band linac is divided in two main sections: the low energy one, composed of $4$ accelerating sections, carries the electron beam up to the maximum energy $E_e = 320\,$MeV. A dogleg transport line downstream its exit, delivers the beam at the Low Energy Interaction Point (LE IP), avoiding in this way the bremsstrahlung radiation contribution. The downstream high energy linac is composed of $8$ accelerating sections and brings the electron beam up to the maximum energy $E_e = 740\,$MeV, then the electron beam, passing through a dogleg beam line, reaches the High Energy Interaction Point (HE IP). In each of the two interaction point regions a quadrupole triplet provides a flexible final focusing for matching the electron beam and the counter-propagating laser pulse spot sizes. 

The GBS relies on three laser-beam systems: the photocathode laser,  and two interaction point laser system. The photocathode laser-beam system is based upon a Ti:Sa chirped pulse amplification system. The photocathode laser-beam system manufactured by Amplitude Technologies~\cite{AmplitudesTechnologies} is composed of a standard $100\,$Hz front-end delivering a $25\,$mJ chirped pulse. The front-end is directly seeded by the optical master oscillator of the synchronization system of the GBS to reduce phase jitter. The laser pulse is then sent to a multipulse cavity generator providing $32\,$ pulses synchronized to the rf. The pulse train is amplified in a multipass Ti:Sa cryogenically-cooled crystal pumped by 12 pump laser-beams that are time delayed. After the optical compressor and third harmonic generator, $32\,$ UV pulses with an energy of a few hundred of microJoules each are transported 15 meters away to the copper photocathode of the GBS. 

Amplitude Syst\`emes~\cite{AmplitudesSystemes} developed Yb:YAG high power picosecond laser-beam systems that employ frequency doubling to produce laser pulses with energies in excess of $0.2\,$J at $515\,$nm central wavelength and with $3.5\,$ps (RMS) pulse duration at a repetition rate of $100\,$Hz. The Yb:KGW femtosecond oscillator is synchronized on the GBS reference clock, pre-amplified and chirped before injecting in a water-cooled thin disk Yb:YAG regenerative amplifier. The output, a $30\,$mJ chirped pulse is amplified in a low-vibration cryogenically-cooled thick disk mutli-pass amplifier up to $350\,$mJ. After the optical compressor, a $515\,$nm, $5\,$ps FWHM duration, $200\,$mJ pulse is generated in a Triborate lithium ($LiB_{3}O_{5}$) second harmonic generator.  

In order to reach the foreseen world-class performance of the GBS, a new and original interaction point (IP) module has been designed. It is based on a passive optical system able to recirculate and focus 32 times a single laser-pulse at the IP. This laser beam circulator (LBC) has to ensure not to spoil the excellent and unprecedented small bandwidth of the $\gamma$-ray beam. To achieve this tight requirement, a constant polar crossing angle $\phi$ between laser and electron beams must be maintained during the whole circulation process. In order to preserve a high number of scattered photons, the excellent overlap of the laser and electron beam distributions must also be preserved while the laser beam is circulated, thus tightly constraining the geometrical superposition of the successive laser-beam focus spots at the IP. The design of the LBC has been reported in Ref.~\cite{Dupraz:2014aa}. An effective gain on the laser power, due to the circulation process, in excess of $20$ is required to reach the $\gamma$-ray flux performances of the ELI-NP-GBS.

After each interaction point, at a distance of $\sim 10\,$m, the $\gamma$-ray beam collimation and characterization line delivering the $\gamma$-ray beam to user's experimental area is composed of collimation system, a $1\,$m thick concrete shielding and a complete characterization system. The collimation system consists of $14$ tungsten slits of independently adjustable aperture placed inside a high vacuum chamber. The whole collimation system is placed on a micrometer $6$ degree of freedom radiation hard motorization positioning system. The collimation system produces the required collimation in the $70-700\,\mu$rad to achieve the $\leq 0.5\%$ bandwidth with the associated constraints on its alignment with respect to the $\gamma$-ray beam. The characterization beam line embeds a Compton spectrometer, a Nuclear resonant scattering system, a $\gamma$-ray beam profile imager and $\gamma$ sampling calorimeter.

\section{\label{sec:IPmodule} Interaction point module}

% list of figures : general view of the IP module (CAD ?) 

\begin{figure*}[!htpb]
\includegraphics[width=14cm]{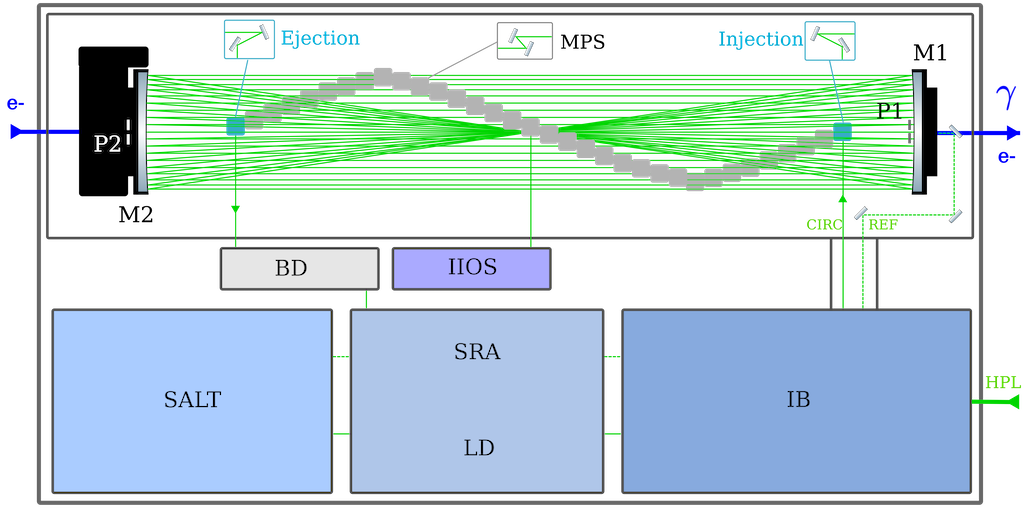}% Here is how to import EPS art
\caption{\label{fig:LBCblocks} Schematic top view of the interaction point module with different sub part SALT: synchronization and alignment laser tool, SRA: self reference alignment,  LD: laser diagnostic, IB: injection box, M1,M2: mirror parabolas, MPS: mirror pair system, and injection et ejection mirror system. The interface with the input high power laser beam (HPL) and electron beam ($e-$) is indicated.}
\end{figure*}

\begin{figure*}[!htpb]
\includegraphics[width=16cm]{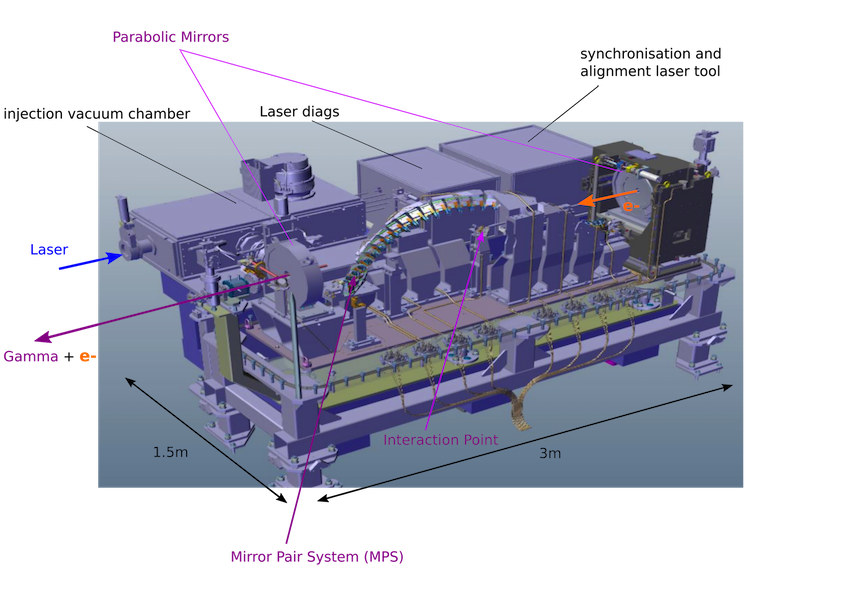}% Here is how to import EPS art
\caption{\label{fig:ipmodule3D} 3D view of the complete interaction point module without the top cover of the LBC vacuum chamber.}
\end{figure*}

The ELI-NP-GBS is constituted of two identical IP modules. Each is composed of the following elements. 
\begin{description}
  \item[Laser beam circulator] it is the core part of the IP module composed of two on-axis parabolic mirrors set in confocal arrangement, 31 mirror pair systems (MPS) each mounted on a piezo-electric rotation stage, a fixed pair of injection mirrors and a fixed pair of ejection mirrors. A piezo-electric motorized inserter placed at the interaction point allows to set various tools on the LBC axis for diagnostics and alignment purposes only. The LBC as a whole is enclosed in an ultra high vacuum (UHV) vessel.
  \item[Synchronization and alignment laser tool] SALT is embedded in the IP module and can be remotely controlled to align and synchronize the LBC and monitor its performance. 
  \item[Injection box]: IB is a high vacuum vessel enclosed optical breadboard with various optics and laser beam diagnostics. It distributes and stabilizes the various laser beams delivered by the SALT and the main high power laser beam for their injection in the LBC. Near and far field laser beam profilers allow to set accurate injection references for the laser injection in the LBC. 
  \item[IP imaging optical system] The IIOS is a high resolution optical imaging system coupled to two detectors, one giving a large field of view at the IP and one giving a time resolution down to $2.7\,$ns. 
  \item[Cavity beam position monitors] CBPMs are installed on the mechanical support fixed to the base Invar\texttrademark plate of the LBC. They provide precise transverse position measurements of the electron bunches at the entrance and exit of the IP module. 
  \item[Laser-beam diagnostics] LD are embedded in the IP module to monitor the injected high power laser beam parameters relevant for the Compton scattering $\gamma$-ray generation and laser beam circulation. Laser-beam power, pulse length, wavefront and polarisation are monitored by the LD.
\end{description}
A schematic view of the interaction point module is given in Fig.~\ref{fig:LBCblocks}. 
  % figure scheme of the ST 
It was taken from the downstream part of the interaction point. The various instruments can be distinguished on the left side, the large LBC vacuum chamber is on the right. The downstream CBPM support is also visible. Another view with more details on the LBC elements and electron and laser beam directions is given in Fig.~\ref{fig:ipmodule3D}. Details of the various components of the interaction point module are presented in the remaining of this section. 
% figure 3D model of the interaction point 

\subsection{\label{subsec:IPparam} Interaction point parameters}
A complete simulation code has been developed \cite{Dupraz:2014aa} to determine the specifications and tolerances allowed on the manufacturing of the optics, the motorization and the opto-mechanical elements. It is based on two different approaches:
\begin{itemize}
	\item a fast computing time ray tracing and $4\times4$ ABCD matrices for geometrical purposes, and in particular alignment \cite{Dupraz:2015aa};
	\item a slow computing time beam synthesis propagation provided by \textbf{Code~V}\cite{CodeV:software} for optical performances studies.
\end{itemize}
These two different calculations are managed by a master code that hold the geometry parameters to ensure a consistency between the two codes.
The essential design parameters and associated tolerances to achieve the required performances of the GBS are summarized in Table~\ref{tab:paramIP}.

\begin{table}[b]
\caption{\label{tab:paramIP} Summary of the optimized parameters to reach ELI-NP-GBS specifications, using design electron beam parameters \cite{Adriani:2014aa}. The reader is referred to Ref. \cite{Dupraz:2014aa} for a  detailed description of the meaning of the parameters.}
\begin{ruledtabular}
\begin{tabular}{lccc}
 \textbf{Parameters} & \textbf{Value} & \textbf{Tol.} & \textbf{Unit} \\
\hline
Interaction & & & \\
\hline
$w_0$, waist 			& 28.3 & $\pm$3 & $\mu$m \\
$\phi$, crossing angle			& 8.00 & $\pm0.03$	 & $^\circ$ \\
$\lambda_0$, central wavelength & 515 & $\pm1$ & nm \\
\hline 
Optics & & & \\
\hline
$F_{max} $, laser fluency		& $>0.4$& - & J/cm$^2$ \footnotemark[1]\\
$D_M$, clear aperture		& $>27$ & - & mm \footnotemark[2]\\	
\hline 
Mirrors pairs system & & & \\
\hline
$D_{MPS}$, inter-mirror distance 		& 40.4 & $\pm0.05$ & mm \\
$\theta_{0}$, AOI		& 23.8 & $\pm0.2$	 & $^\circ$ \\
$|\Delta l_{max}|$, path length tuning			& $2.53$ & - & mm \\
$\epsilon$, mirror parallelism 		& $0$ & $<4$ & $\mu$rad \\
\hline
Laser beam circulation & & & \\
\hline
$N_{pass}$, circulation number 		& 32 & - &- \\
$l_c=46\cdot c/\nu_{rf}$ & 4828.59 & $\pm 0.003$ & mm \\
$D$, distance between parabolsa 				& 2377.31 & $\pm0.05$ & mm \\
$R_C$, corona radius  			& 166 & $\pm1$ & mm \\
$w_M$, input laser beam waist			& 8.3 & $\pm 0.2$ & mm \\
\hline
Performances & & & \\
\hline
TASD 			& 41000 & - & $\gamma/(s.eV)$  \footnotemark[3]
\end{tabular}
\end{ruledtabular}
\footnotetext[1]{Laser induced damage threshold required for,$E_L=400\,$mJ, the IP laser beam energy at the high-energy interaction point of the ELINP GBS }
\footnotetext[2]{Minimal pupil diameter}
\footnotetext[3]{Maximum time averaged spectral density calculated for a perfect system at the low-energy interaction point of ELINP-GBS at $E_\gamma = 2\,$MeV with $E_L=200\,$mJ, and $Q=250\,$pC}
\end{table}

\subsection{\label{subsec:IPmechanicalFrame} Interaction point module mechanical frame}

\begin{figure}[!htpb]
\includegraphics[width=8cm]{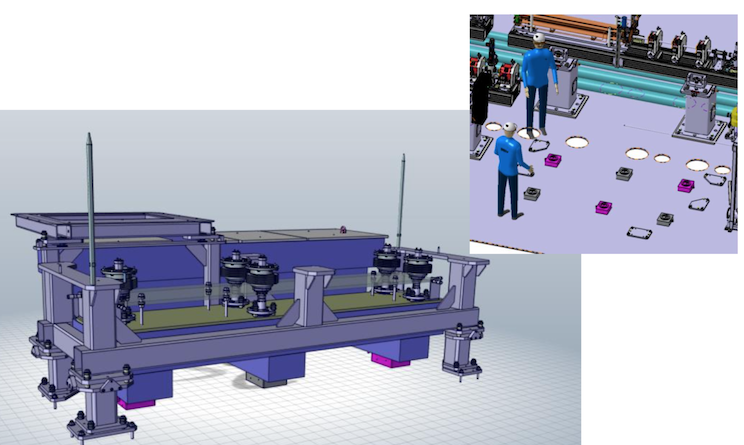}% Here is how to import EPS art
\caption{\label{fig:ip-mechanicalframe} 3D view of the mechanical frame. The intermediate table can is semi-transparent. The INVAR base table is the yellow plate. The internal optical breadboard sits on the top of the 5 feets.}
\end{figure}

The whole IP module is set on a granite structure providing a thermal inertia. The granite structure is mounted on six, $3-$axis adjustable, feet for initial alignment in the mechanical reference network, see Fig.~\ref{fig:ip-mechanicalframe}.  A mechanical frame independent of the granite block holds the LBC and the IB vacuum vessels to decouple the vacuum chambers and the inner optical breadboards. The LBC optical table is supported by 5 feet, screwed on an Invar\texttrademark \footnote{Generally known as FeNi36, common grade of Invar\texttrademark has thermal expansion coefficient of $1.2\times10^{-6}m/K$} plate that is set on the granite block. This intermediate optical table can be moved from air to adjust the whole inner-vacuum optical breadboard in order to extend the final alignment range of the laser focal spot on the electron beam. The SALT and LD can be removed by the help of a mobile crane and placed back into position thanks to a set of precise localization pins.

% figure 3D model of the interaction point 

\subsection{\label{subsec:introduction} Laser beam circulator}
The LBC is made of three groups of optics: the two parabolic mirrors M1 and M2, an injection mirror-pair M$0$-M$0'$ and an ejection mirror-pair  M$33$-M$33'$ and the $31$ MPS (Fig. \ref{fig:LBCblocks}). The parabolic mirror M2 is mounted on a five degree of freedom motorized mount suitable for UHV operations. Each MPS is mounted on a piezo-electric rotation stage. 

% figure 3D CAD model / top view  

\subsubsection{Parabolic mirrors}
\begin{figure}[!htpb]
\includegraphics[width=8cm]{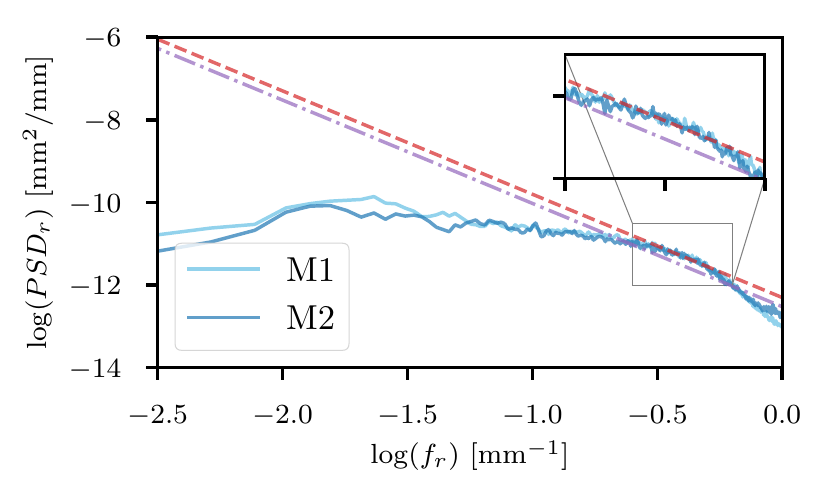}% Here is how to import EPS art
\caption{\label{fig:parabola-PSD} Radial PSD of the (light blue) M1 and (dark blue) M2 parabolic mirrors measured over the useful area with the specified (red-dashed) upper-limit and (violet-dash-dotted) lower-limit lines. Courtesy of SAFRAN-REOSC }
\end{figure}

Parabolic mirrors specifications have been drawn based on simulations using a model of the system~\cite{Dupraz:2015ab}. The mirrors have been polished and coated by REOS-SAFRAN~\cite{REOSC} with the characteristics gathered in Table.~\ref{tab:table2}. The substrates are made of high grade Fused Silica. The optics cosmetics are MIL-Prf-13830B~\cite{MILF} (S/D 20-10) corresponding to high-power laser-beam grade optics. Careful attention must be paid to the mid-spatial frequencies of the power spectrum density (PSD) of surface defects, particularly to the PSD gradient in the useful corona (see Fig. \ref{fig:LBC}) which is leading to a spread of the focii points of the various laser-beam passes as they are reflected at different azimuthal position by the useful corona of the parabolic mirrors. This is one of the most stringent specification of the parabolic mirrors. The targeted surface characteristics have been achieved by the polisher and the measured PSD is shown on Fig.~\ref{fig:parabola-PSD} for the two parabolas along with the design specification.

\begin{table*}
\caption{\label{tab:table2} Polished and coated parabolic mirrors optical parameters inferred from measurements of the parabolic mirror prior coating and on witnesses for coating properties}
\begin{ruledtabular}
\begin{tabular}{llcccl}
 Parameters &  description & requirement & meas. value & error & unit \\
\hline
polishing & & & & & \\
\hline
$ROC_1$;$ROC_2$ & M1;M2 radius of curvature	& $2377.31$	& $2377.240$;$2377.585$  & $\pm0.04$ & mm\\
$ROC_1/ROC_2$ & radii of curvature ratio	& $1$ & $1$ & $\pm0.001$ & - \\
$r_{max};r_{min}$ & useful area corona radii &  -& $138;194$ & $\pm 0.5$& mm \\
$PSD_r^{(1)}$ & surface irregularity $0.0027\leq f_r \leq 1\,$mm$^{-1}$ & $K/f_r^\alpha$ & fig. \ref{fig:parabola-PSD} &  \footnotemark[1]  \footnotemark[2]  & \\
$PSD_r^{(2)}$ & surface irregularity $f_r>1\,$mm$^{-1}$ & $K/f_r^\beta$ &  fig. \ref{fig:parabola-PSD} &  \footnotemark[2] & \\
$\sigma_1$;$\sigma_2$ & M1;M2 surface roughness  &$<1$ & $0.536$;$0.566$ & 	0.1 & nm \\
\hline
coating & & & & &  \\
\hline
$\lambda_0$ & central wavelength	& 515 & $515$ & $\pm1$ & nm \\
$AOI$& angle of incidence	& 4 & $4$& -  & $^\circ$ \\
$R$ & Reflectivity 	& & $>99.85$ & $0.05$ & $\%$ \\
$|R_s-R_p|$ & difference between $S$ and $P$ polarization reflectivities & $<500$ &  $<250$ & - & ppm \\
$LIDT$ & for pulses $515\,$nm,$3.5\,$ps, $100\,$Hz, $3000h$ 	& $>0.4$	& $>0.5$ & -  & J/cm$^2$\\
$C_i$ & vertex position accuracy & & - & $\pm1$ & mm \\
 & vertex position measurement accuracy & & &$\pm0.5$ & mm \\
\end{tabular}
\end{ruledtabular}
\footnotetext[1]{$3\times 10^{-13} \leq K \leq 5\times 10^{-13}$, $\alpha \geq  2.5$},see curves in figures. 
\footnotetext[2]{$\beta \geq 2$, see curves in figures, integral of the PSD in the useful area $\sigma_1 = 4,7\,$nm,$\sigma_2= 5.\,$nm, measured with IDEOS interferometer}
\footnotetext[3]{Average of the measurement made on 3 points along the radius in the useful corona, $r_1=138\,$mm,$r_2=166\,$mm and $r_3=194\,$mm}
\footnotetext[3]{LIDT : laser induced damage threshold measured by LIDARIS\cite{Lidaris} for $3.5\,$ps (RMS) pulse length $515\,$nm, central wavelength laser under vacuum at  AOI$=4^\circ$ and at $f_{rep}=50\,$kHz for $10^{7}$ shots}
\end{table*}

% fig PSD parabola. 

The parabolic mirror M1 is mounted on a manually adjustable five-axis mount thanks to a series of high turn per inch screws to get enough accuracy for the tilt setting. The typical resolution obtained with a $1^\circ$ screw rotation is of $20\,\mu$rad. This requirement is driven by the fact that the alignment of M1 with respect to the global GBS mechanical reference frame defines the final orientation of the LBC axis. The corresponding error budget for the M1 alignment in the GBS mechanical reference frame is of $\pm 50\,\mu$m, $\pm 100\,\mu$rad.

\begin{figure}[!htpb]
\includegraphics[width=8cm]{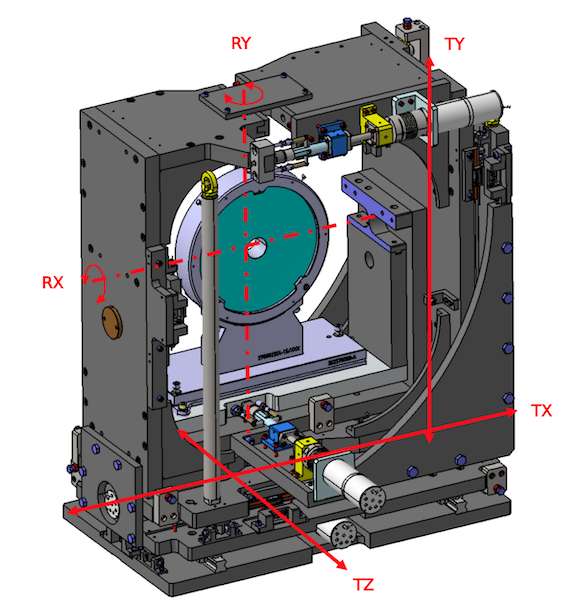}% Here is how to import EPS art
\caption{\label{fig:lbc-5ddl} 3D view of the mechanical model of the five-axis motorized UHV compatible mount of M2.}
\end{figure}

The five-axis motorized mount of M2 has been realized by the IRELEC-ALCEN company~\cite{Irelec}. A 3D view of it is given in Fig.~\ref{fig:lbc-5ddl}. It is based on two large $TX$ and $TZ$ axes tables. The vertical $TY$-axis is set on a shelf bracket on top of the $TX$ moving part. The $TY$ moving part holds a two rotations flex frame for the $RX$ and $RY$ motions. The characteristics of the five-axis motorized M2 mirror mount are collected in the table Table~\ref{tab:table3}. Each axis is equipped with two radiation hard, UHV compatible, high accuracy compact electrical limit-switch (SW-3 type) \cite{Irelec}, a high resolution optical encoder and a reference mark. The axes are driven by a Phymotion\texttrademark\ controller~\cite{Phymotion}. 
 \begin{table}[!htpb]
\caption{\label{tab:table3} The performance of the five-axis motorized mount loaded with M2 and the related mechanical frame.}
\begin{ruledtabular}
\begin{tabular}{llcc}
Axis & parameters & value & unit \\
\hline
$TX$, $TY$, $TZ$  & & & \\
  & resolution		& 0.23  &  $\mu$m \\
   & precision 		& 0.1 &  $\mu$m \\
   & repeatability         & 0.1 &  $\mu$m \\
   & travel range        & $\pm3$ & mm \\
   & travel speed      & $\leq 0.5$ & mm/s \\
   
$RX$ (pitch) & & &  \\
  & resolution		& 0.33  &  $\mu$rad \\
   & precision 		& 0.33 &  $\mu$rad \\
   & repeatability         & 0.33 &  $\mu$rad  \\
   & stroke                & $\pm2.5$ & $^\circ$ \\
    & rotation speed      & $\leq 0.5$ & mrad/s \\
    
$RY$ (yaw)		&	&  &    \\
  & resolution		& 0.36  &  $\mu$rad \\
   & precision 		& 0.14 &  $\mu$rad \\
   & repeatability         & 0.14 &  $\mu$rad  \\
    & stroke                & $\pm2.5$ & $^\circ$ \\
    & rotation speed      & $\leq 0.5$ & mrad/s \\

\end{tabular}
\end{ruledtabular}
\end{table}

\subsubsection{Mirror Pair System} 
The MPS is one of the key-elements of the LBC since it allows the circulation of the laser-beam pulses 32 times at the IP. Each MPS is meant to be an optical invariant so that the laser-beam exits the system parallel to the input. Tight mechanical requirements are imposed on this system in order to avoid vignetting of  the laser-beam and marring the alignement of the LBC \cite{Dupraz:2014aa}. The MPS mirrors parallelism must stand a mechanical shocks that can be generated during the installation in the LBC, pressure variation from atmospheric to ultra high vacuum pressure, and have elastic behavior to thermal stress of one Kelvin

\begin{figure}[!htpb]
\includegraphics[width=8cm]{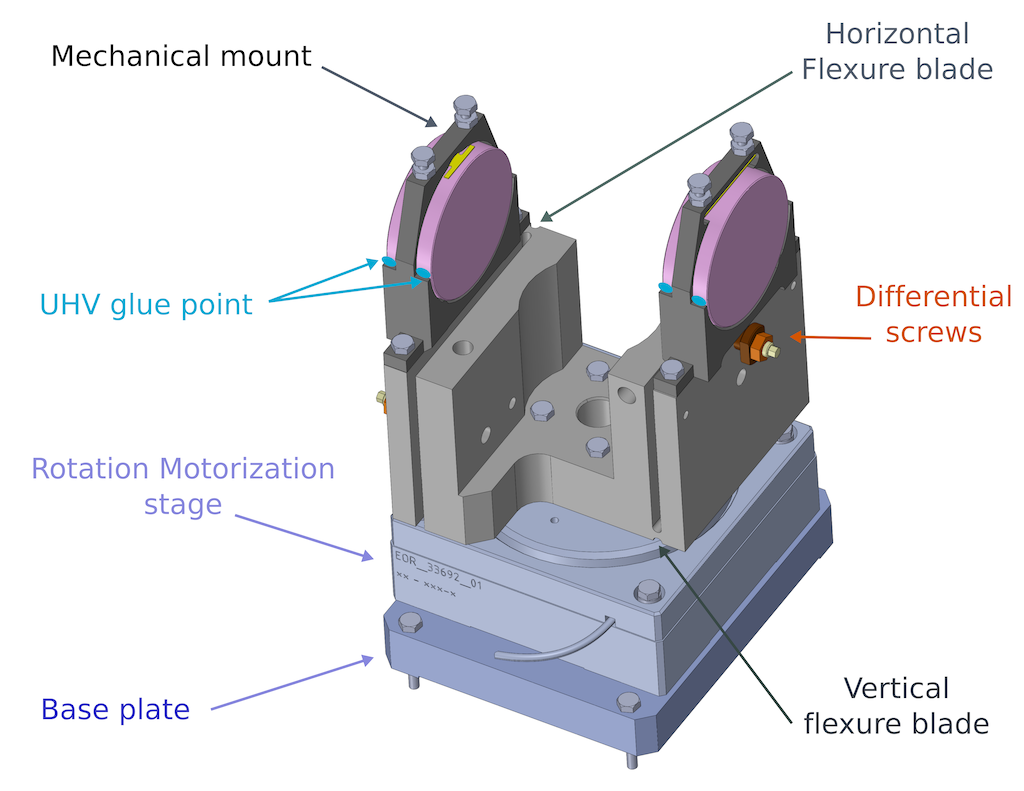}% Here is how to import EPS art
\caption{\label{fig:MPS-3D} 3D view of the complete MPS system}
\end{figure}

Two technological solutions were studied. The first one is based on a flexor to set the angles between the two mirrors maintained within flanges by a lateral-side fixation-retaining leaf-spring. In the second one the mirrors are glued on their fixation holder with an angular tuning control on one of the mirrors while the reticulation process takes place. The first solution has the main advantage of an angular tuning but, in the meantime, presents a worse mechanical stability due to the presence of the leaf spring and the flexor differential screws. The second solution has very good mechanical stability but implies a complicated process to control the parallelism during the gluing process.  An hybrid solution has been integrated, in which the mirrors are glued in the mirror flange. The realization of this last solution is shown in Fig.~\ref{fig:MPS-3D}. 
In detail, each MPS is composed of two flat high grade mirrors, two mechanical mirror mounts, a two-blades flexor, a piezo-electric motorized rotation stage and a positioning base plate.  The mechanical mirror mount is made of stainless steel. The mirror mount is sharpened at the equator to avoid beam vignetting. The mirror sits on three pins on the bottom of the mirror mount. The mirror position can be secured by a $0.25\,$mm leaf spring bent on the top of the mirrors by two screws.  The two-blades stainless-steel flexor is bent thanks to a differential screw made of brass. The inner screw is screwed in a swiveling nut. 

The  motorization of the MPS allows a rotation of the mirrors pairs around $\theta_0$ within a range of $\pm2.5^\circ$ to allow for the synchronization of each pass of the LBC with the electron bunch train. It is a modified version of piezo-electric rotator ECR3030-UHV from Attocube GmbH~\cite{Attocube}. It has a torque of $2\,$N/cm with a bi-directional repeatability of $17.6\,\mu$rad and a sensor resolution of $0.2\,\mu$rad. A step of $\delta \theta$ on the rotation gives a time delay on the circulation period of  $\delta t = \Delta l /c = 2 D_{MPS} (\cos(\theta_0)-\cos(\theta_0-\delta \theta)) /c $ yielding a minimal time step for the synchronization $\delta t_{min} \approx 20\,$fs. The rotation stage is attached to the positioning base plate, equipped with a precision pin locator. The MPS base plate permits the fixation of the whole MPS on its stand in the LBC.  

\begin{figure}[!htpb]
\includegraphics[width=8cm]{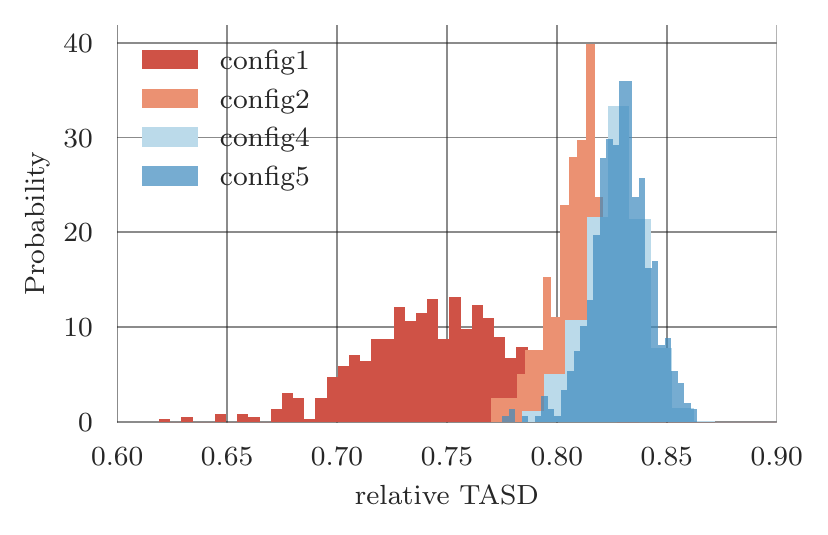}% Here is how to import EPS art
\caption{\label{fig:MPS-ordering} Distribution of the relative time average spectral density of $\gamma$-ray for the various simulated ordering configuration 1, 2, 4, and 5}
\end{figure}

The MPS mirrors have been polished by ELDIM~\cite{ELDIM} by combining Magneto Rheological Finishing (MRF)~\cite{QEDMRF} polishing technology and fine and accurate hand polishing to randomize and reduce the typical micro-structures of MRF polishing, thus limiting any systematic surface error. Their characteristics are shown in Table~\ref{tab:MPSoptics}. They are made of fused silica with high-grade cosmetic aspect (S/D 10-5). 
 \begin{table}[!htpb]
\caption{\label{tab:MPSoptics} Measured and specified optical properties of the MPS mirrors.}
\begin{ruledtabular}
\begin{tabular}{llll}
parameters & requirement & measurement  & unit \\
\hline
optics & & & \\
\hline
surface quality, PV			& $<\lambda_0/20$ 	& $<26$	& nm\footnotemark[1]\\
$<|Z_2^{0}|>$ average focus  	& $ 0$			& $1.1$ 	& nm\footnotemark[2]\\
$|Z_2^{0}|$ max focus         	 & $10$ 			&  $10$ 	& nm \\
\hline
coating & &  &  \\
\hline
Reflectivity				& $>99.95$ 		&  $>99.965$  	&  - \\
AOI						& 23.8 			&  $23.8$		&  $^\circ$ \\
LIDT 					& $>0.4$ 			&	$>0.8$	& J/cm$^2$ \footnotemark[3]\\\
$|R_s-R_p|$ 				& $<500$ 			&	$<100$	& ppm \\
environment 				& $<5\times 10^{-8}$ &  -  &  mbar \\
\end{tabular}
\end{ruledtabular}
\footnotetext[1]{For $\lambda_0=515\,$nm, pupil $26\,$mm-diameter}
\footnotetext[2]{See reference \cite{Wang:1980aa}}
\footnotetext[3]{LIDT: laser induced damage threshold measured by LIDARIS for $3.5\,$ps length $515\,$nm central wavelength laser under vacuum at $23^\circ$ AOI at $f_{rep}=50\,$kHz for $10^{7}$-on-1}
\end{table}
It has been shown that systematic macroscopic surface error on the MPS mirror significantly limits the LBC performances \cite{Dupraz:2015ab}. In the pursue of reducing this defect, Institut Fresnel developped a process \cite{Lumeau:2017aa} to compensate for the coating stress and also to randomize the focus surface errors corresponding to the Zernike coefficient $Z_2^0$ in the Zernike decomposition of the mirror surface~\cite{Wang:1980aa}. Each substrate has been controlled with NewView Zygo\texttrademark\ 7300~\cite{ZygoNewview}. The back-coating compensation is then calculated after calibration studies have been made~\cite{Lumeau:2017aa}. The mirrors have been coated with a Leyblod\texttrademark Optics Helios~\cite{Leybold, Zoller:1994aa} at Institut Fresnel. The multi-layer coatings are constituted of Hf$_2$O$_5$ high index and SiO$_2$ low index layers. The multi-layer stacking has been optimized to get a high-reflectivity and small difference between the s-polarized R$_s$ and p-polarized R$_p$ laser-beam reflectivities.
\begin{figure}[!htpb]
\includegraphics[width=8cm]{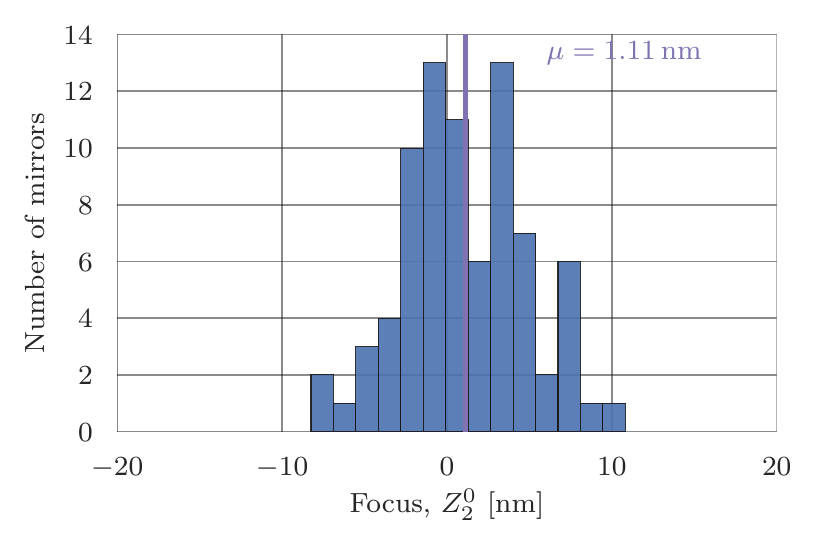}\\% Here is how to import EPS art
\includegraphics[width=8cm]{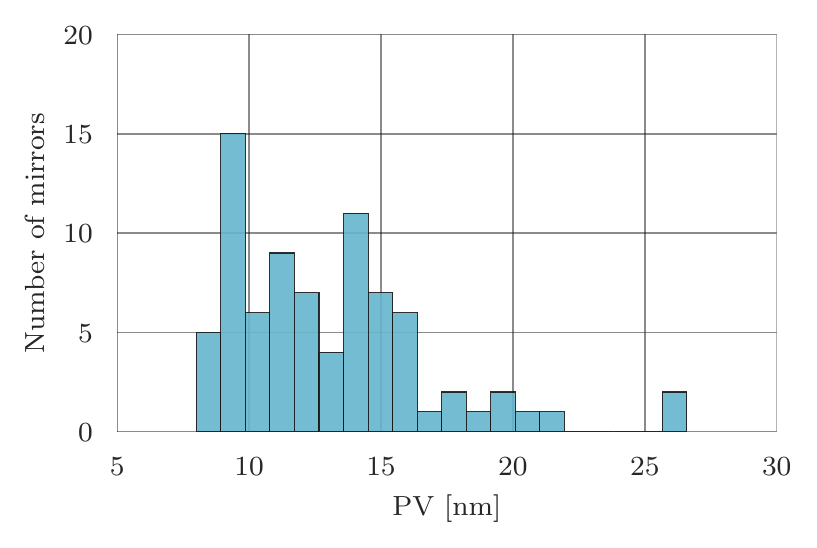}% Here is how to import EPS art
\caption{\label{fig:MPSoptics-PVfocus}  (Top) distribution of the Zernike Coefficient $Z^0_2$ (focus) for the 82 mirrors coated at Institut Fresnel and controlled on a pupil of 29 mm and a Zernike decomposition realized on a 26 mm diameter pupil. (Bottom) Distribution of the 82 mirrors PV surface errors measured on a 26mm diameter pupil.}
\end{figure}  
The mirror's active surfaces have been controlled again after the coating took place. The measured distribution of Peak-to-Valley (PV) surface errors and $Z^0_2$ focus Zernike amplitude are shown in Fig.~\ref{fig:MPSoptics-PVfocus}. It has been decided to polish and coat $80$ mirrors in order to have the possibility to select the best combination of 62 mirrors. In order to reduce the cumulated focus while also preserving a low PV wavefront error within the Marechal criterion for all passes~\cite{Smith:2007aa}, an algorithmic ordering procedure of the MPS mirrors has been applied~\cite{Ndiaye:2019aa}.  Using the developed simulation code several simulations with different configurations have been run. The real surface maps of the coated parabolic and MPS mirrors have been implemented for these simulations. In all the simulations the injection mirrors are considered perfect, meaning that the residual focusing error in the wavefront is compensated in the laser system and laser transport beam line up to the first pass and that the parabolic mirrors are confocal and perfectly aligned. For each simulation, the orientation of the surface maps to be attached to each mirror is randomly selected between $0$ and $2\pi$ and the maximum parallelism defect of the MPS mirrors is fixed to $\Delta \epsilon = 7.5~\mathrm{\mu rad}$.  From one simulation to another for a given configuration, both the mirrors and their order are identical.  Various selection rules configuration have been tested. The results are displayed in Fig.~\ref{fig:MPS-ordering}. The configuration 1 corresponds to a random selection of the mirrors without ordering. The configuration 2 consists in selecting the 62 mirrors by placing the mirror with the best focus and PV in first position and sequentially minimizing the absolute value of the accumulated focus for the following mirrors. 
For the configuration 4, the 80 mirrors are classified by the ordering algorithm in which the absolute value of the accumulated focus is constrained not to exceed $f_{thresh} = 2\,$nm at each pass. 
 When the ordering algorithm does not succeed to find a mirror allowing to keep the cumulated focus below $f_{thresh}$, the best PV mirror is automatically selected.
The configuration 5 is identical to the 4 except that $f_{thresh} = 1.5\,$nm. When the algorithm does not find a mirror candidate to keep the cumulated value of the focus below $f_{thresh}$, the threshold value is incremented to $f_{thresh}+0.5\,$nm. When a mirror candidate is found, the $f_{thresh}$ is reset to the initial value of $1.5\,$nm for the next mirror. 

It is observed in Fig.~\ref{fig:MPS-ordering} that the relative TASD improves by $0.06$ in configuration 2. The configuration 5 is the best, the relative TASD improving by $\sim0.09$  and is the chosen classification scheme. 

\begin{figure}[!htpb]
\includegraphics[width=8cm]{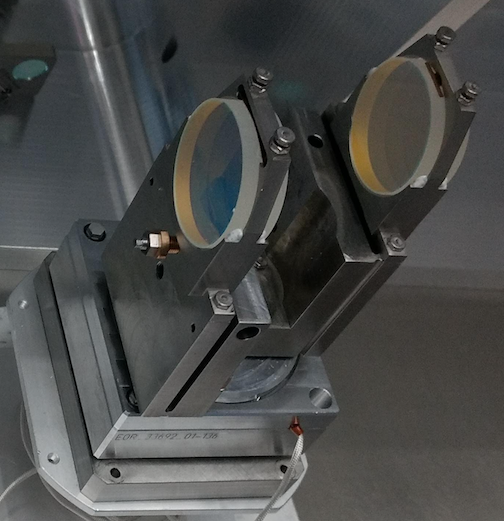}% Here is how to import EPS art
\caption{\label{fig:MPS-glue} Picture of a MPS ready to be installed under vacuum immediately after glueing, see text for details on the procedure.}
\end{figure}

The inter-mirror distance $D_{MPS}$ is first set with a PEEK \footnote{Polyether ether ketone (PEEK) is a colourless organic thermoplastic polymer in the polyaryletherketone (PAEK)} spacer and a rough parallelism steering is then realized. The mirrors are then glued with an epoxy two-parts Varian TorrSeal\texttrademark\ ~\cite{Torrseal} low outgassing UHV-compatible glue with high radiation hardness. Several tests have been done to get the most stable and simple glueing process. Four points of glue are applied, as shown on Fig.~\ref{fig:MPS-glue}, with a calibrated needle to minimize the amount of glue and to get a reproducible application of its quantity. They are placed at the equatorial position of the mirrors in the front and back part of the mirror mechanical mount. Immediately after application the MPS are put in an evacuated chamber with a pressure below $1.5\times10^{-1}\,$mbar for at least $24\,$hours while the reticulation process takes place.
\begin{figure}[!htpb]
\includegraphics[width=8cm]{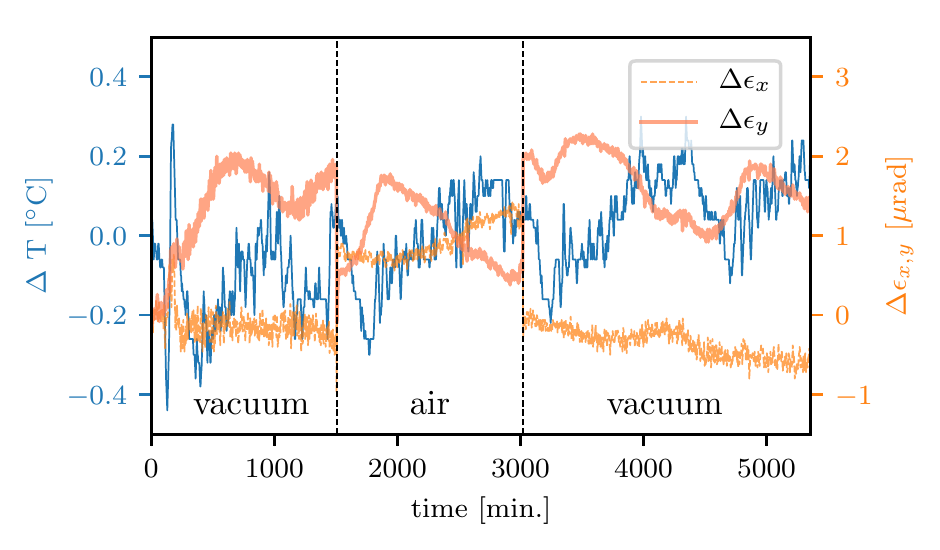}% Here is how to import EPS art
\caption{\label{fig:MPS-vac1} MPS parallelism stability under vacuum while the air temperature is monitored for about 3.5 days. This measurement is performed after glue reticulation for 24 hours under vacuum followed by a reticulation time of 24 hours at air and then vacuum again.}
\end{figure} 
The angular beam deviation induced by the MPS parallelism instability is measured with an electronic autocollimator, UltraSpec\texttrademark\ TA US 300-57, with a focal length $f=300\,$mm \cite{TriopticsUltraSpec}. It has been modified by a stainless steel support to improve its vertical stability. 

The stability of the MPS system has been investigated in air and vacuum over a long time period. A double self-referencing measurement is performed by use of a reference mirror and the reflection on the back-surface of the first mirror of the MPS. The measurement is made with the MPS placed such that the laser-beam of the autocollimator impinges the MPS close to normal incidence. The accuracy and precision of the autocollimator are $\pm 1.2\,\mu$rad and $\pm 0.02\,\mu$rad, respectively. A long term measurement is shown on Fig.~\ref{fig:MPS-vac1}. It is observed that the horizontal angle beam deviation $\Delta \epsilon_x$ varies slowly with time while the vertical angle beam deviation $\Delta \epsilon_y$ is fluctuating faster and follows a daytime period roughly anti-correlated with the air temperature variation. We can notice a little variation in the $x$ and $y$ in the parallelism at each transition air-vacuum and vacuum-air. The amplitude of the variation is small $\Delta \epsilon_x^{vac-air} \sim -1.5\,\mu$rad and $\Delta  \epsilon_y^{vac-air} \sim +1.8\,\mu$rad. 

The mechanical stability of one of the MPS has been checked by repeating the procedure of installing it inside the LBC several times on localization pins. This operation is delicate, particularly since the insertion of the base support of the MPS can generate a mechanical shock of a few g. The variation of the angular beam deviation  is measured by the autocollimator to be within the specification of $\Delta \epsilon= \sqrt{( \Delta \epsilon_x^2+ \Delta \epsilon_y^2)}<4\,\mu$rad for a single MPS.

In order to maintain an excellent MPS parallelism, and not to spoil the LBC performance, the angular beam deviation measured by the autocollimator is specified to be within $4\,\mu$rad. 

\begin{figure}[!htpb]
\includegraphics[width=8cm]{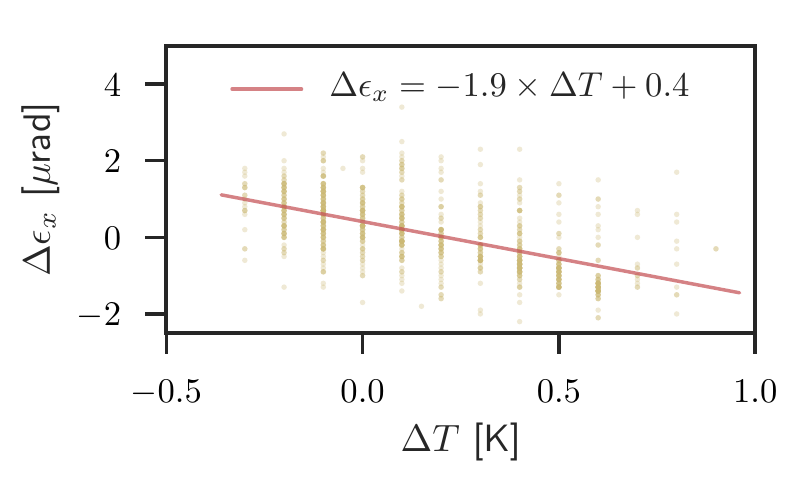}\\% Here is how to import EPS art
\includegraphics[width=8cm]{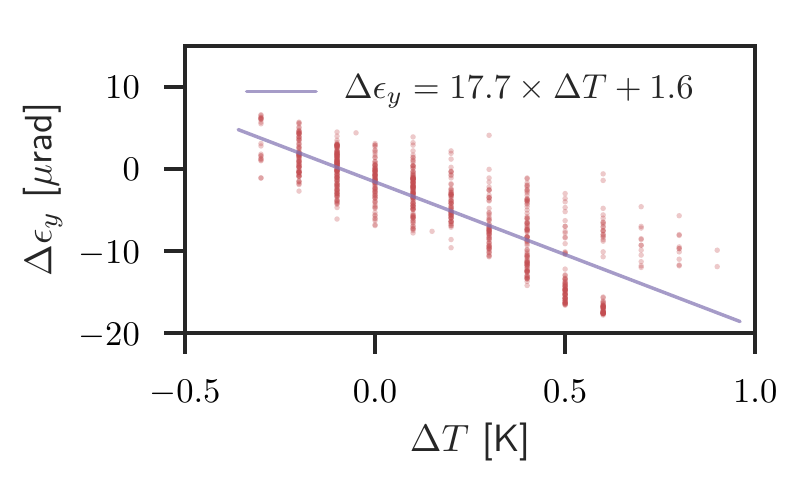}% Here is how to import EPS art
\caption{\label{fig:MPS-temp} MPS beam angle deviation as function of the air temperature measured near the body of the MPS. The working temperature was $24^\circ$C.}
\end{figure}

The parallelism is affected by the thermal dilatation/contraction of the different materials of the MPS opto-mechanics.
The thermal sensitivity of the MPS parallelism has been studied at various working temperatures between $20^\circ\,$C and $25^\circ\,$C. The foreseen temperature in the accelerator tunnel of ELI-NP-GBS is $22\,\pm0.5^\circ\,$C. Fig.\ref{fig:MPS-temp} shows the angular beam deviation on both axis as function of the temperature. 
The thermal sensitivity is found different for the two axes due to the shape of the mechanical design. The thermal sensitivity of the vertical and horizontal parallelism is $a_y = -17\pm5\,\mu$rad/K and $a_x = - 1.9\pm1\,\mu$rad/K, respectively.  
The deviation to perfect parallelism is significantly larger than specified. It appears that the vertical axis is most sensitive to temperature variation. The horizontal thermal angular drift is very low and compliant with the requirement.  As a result, a thermal stabilization of the MPS installed in the LBC is required. A UHV compatible ceramic platinum thermo-resistance (PT100, class $1/3$ C) cabled with four Kapton insulated wires has been implemented to monitor the temperature of the mount.  The probe is fixed with silver UHV conducting glue on a small piece of stainless steel and then fixed on the body of the MPS flexor thanks to one of the screws used to fix it on the rotation stage. Four MPS have been equipped with such thermometers. A picoLog\texttrademark\ PT104 four-channel temperature data logger is used. A high-performance 24-bit ADC is employed to achieve a $0.001^\circ\,$K resolution and an accuracy of $0.015^\circ\,$K.  These thermometers provide precise and under vacuum monitoring of the MPS temperature that is critical to monitor the LBC performance.

An important discrepancy has showed up during the tests between the differential screws inducing a coupling between the $x$ and $y$ axes. This default is one of the main mechanical limitations of the present design that complicates greatly the parallelism tuning. For some MPS a mechanical relaxation of the ensemble of the flexure blade and the differential screw has been observed as long as a few hours after the setting. Consequently, those MPS had to be reset. 

\subsubsection{Injection and ejection mirrors }

The optics used for the injection and ejection mirror systems, see Fig.~\ref{fig:LBCblocks}, have the same characteristics as the MPS optics, that are detailed in Table~\ref{tab:MPSoptics}, except that the M$0'$ diameter is $50.8\,$mm. The injection mirrors, dubbed $M0$ and $M0’$, act like folding mirrors forwarding the laser-beam coming from the IB to the fixed parabolic mirror M1. Due to space constraints a two-mirrors setup has been chosen. 

%\begin{figure}[!htpb]
%\%includegraphics[width=8cm]{fig/fig-M0M0p-3D.png}% Here is how to import EPS art
%\caption{\label{fig:INJ-EJECT} [FIGURE TO BE REDRAWN AND UPDATE no piezo driven actuator.  a - view of the integration of the injection mirrors system allowing an injection in the bottom of the M1 parabola. b - 3D view of the injection (or ejection) mirrors system}
%\end{figure} 

A non vignetting mirror $M0$ (with the reduced dimensions of the MPS mirror) is set within the LBC circulation corona allowing the first mirror to work at $45^\circ$ of incidence. Thanks to this arrangement, the laser-beam is injected in the LBC perpendicularly to the electron beam axis, thus simplifying the vacuum vessel design.
A flexor-based design has been made to allow for a tuning of the position of $M0'$ according to the fixed $M0$ position and the injected laser-beam, as can be seen on Fig.~\ref{fig:LBCblocks}.
The flex brackets are designed with fine, high turn per inch, threaded adjusting screws. The ejection mirrors system is similar to the injection one and allows to steer the laser beam out of the LBC vacuum chamber on a high-power air cooled beam dump and a series of laser diagnostics.

\subsection{Embedded instruments}

In addition to the LBC, the IP module is equipped with three other mains instruments. They are remotely controlled and monitored through a supervision software and a data acquisition system developed by SCARELL\cite{Scarell} for expert diagnostics, initial alignment and synchronisation of the system. It is interfaced with the main EPICS control system of ELI-NP-GBS for its regular operations.

\subsubsection{Synchronization and alignment laser-beam tool} 

\begin{figure}[!htpb]
\includegraphics[width=8cm]{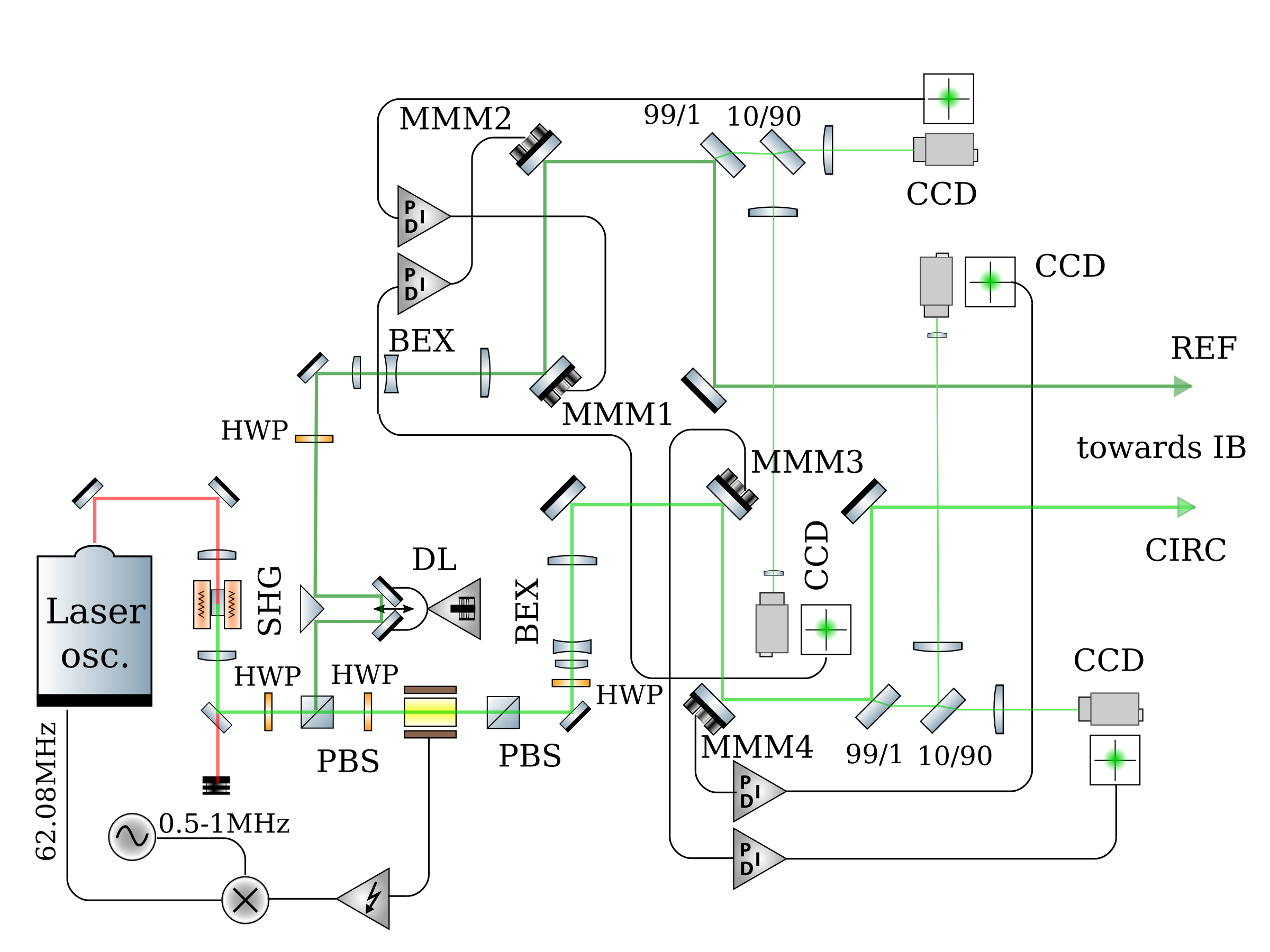}% Here is how to import EPS art
\caption{\label{fig:ST} Optical scheme of the synchronization and alignment laser-beam tool. SHG: second harmonic generator; HWP: half wave retardation plate; DL: piezo-electric motorized delay line; PC: Pockels cell with half wave phase retardation. BEX: beam expander; MMMX: piezo-electric motorized steering mirror mount; CCD: charge coupled device based area detector allowing for laser-beam spot centroid determination; PID: Proportional Integrate Derivative software controllers for the closed loop control of the beams positions and pointings.}
\end{figure}

The SALT, schematized on Fig.~\ref{fig:ST},  uses a low frequency-noise laser-oscillator with a repetition frequency $\nu^{(46)}= \nu_{rf}/46 = 62.08695$ MHz with a  $1030\,$nm central wavelength and bandwidth of  $\approx 13\,$nm (FWHM), delivering an output power in excess of $100\,$mW. It is frequency locked on the optical master clock of ELI-NP-GBS that is distributed by a non stabilized fiber link \cite{Piersanti:2018aa}. The laser pulse is sent in a non-critical phase-matched secondary harmonic generator (SHG) based on a $7$~mm long LBO crystal hold in a temperature-controlled oven that is set at $194^{\circ}$C. The conversion efficiency is $18\,\%$ for standard operations. The measured laser pulse duration after the SHG is $t_p<200\,$fs. The laser pulse is further split in two beam arms to generate a \textit{ reference} beam (REF), used as reference to synchronize the LBC on $\nu^{(46)}$, and a \textit{circulation} beam (CIRC), used for alignment and synchronization purposes, as if it were the main laser pulse that circulates in the LBC under normal operations. The repetition frequency of the CIRC beam is reduced to $700\,$kHz in order to inject only one laser pulse per circulation time period $T_{\textrm{circ},32} = 515.4\,$ns where $T_{circ,n} = n/\nu^{46}$. It is done thanks to a two BBO-crystal Pockels-cell pulse-picker. The measured pulse picking extinction ratio is greater than $1/500$. The REF beam is sent to a piezo-electric motorized delay-line that allows to synchronize the REF beam with the first pass of the CIRC beam in the LBC by remotely tuning the delay over a range of $166\,$ps. As shown, in figure Fig.~\ref{fig:ST},  both REF and CIRC beams are then expanded through telescopes to match the required laser beam size $w_0 \approx 8\,$mm at the LBC input. 
  
  % figure scheme of the ST 

An active-stabilization loop corrects the laser beams low-frequency pointing drifts $du_{x}, du_{y}$ down to $4\,\mu$rad and position drifts $dx,dy$ below $90\mu$m by means of a software PID and motorized mounts.  At the output of the SALT, the CIRC and REF beam powers are $\sim 170\,\mu$W at $700\,$kHz and $1.8\,$mW at $62\,$MHz, respectively. Theses values correspond to a splitting balance of $90\%/10\%$ between the REF and CIRC beams.  These parameters are summed up in Table~\ref{tab:SALT}.
 
 \begin{table}[!htpb]
\caption{\label{tab:SALT} Parameters of the synchronization and alignment laser-beam tool, see text for details.}
\begin{ruledtabular}
\begin{tabular}{llll}
parameters &  REF beam & CIRC beam & unit \\
\hline
$\lambda_0$	& $515$ 	 &	 $515$	& nm \\
$f_{rep}$ 			& $ 62.08$ &  	$0.7$	& MHz  \\
$P$          			&  $1.8$ 		 &  	$0.17$	&	mW \footnotemark[1]\\
$t_p$			&  $<200$		& 	 $<200$		& fs \\
$w_0$			&  $8\pm1$		& 	 $7\pm0.3$		& mm  \footnotemark[2] \\
$dx$				&  $0.09$		& 	 $0.09$		& mm \footnotemark[3] \\
$dy$				&  $0.09$		& 	 $0.09$		& mm \\
$du_x$			&  $3.5$		& 	 $3.5$		& $\mu$rad \footnotemark[4] \\
$du_y$			&  $4$		& 	 $4$			& $\mu$rad \\
\end{tabular}
\end{ruledtabular}
\footnotetext[1]{For a CIRC/REF splitting ratio of 0.9/0.1.}
\footnotetext[2]{Divergence and size can be tuned thanks to 3-lenses beam expander.}
\footnotetext[3]{Active position regulation loop set at $50\,\mu$m threshold with $20\,\mu$m precision on position. The regulation rate is $0.5\,$Hz. RMS values measured with sampling rate of $5\,$Hz}
\footnotetext[4]{Active angle regulation loop set at $3\,\mu$rad threshold with $1.9\,\mu$rad precision on the angle. The regulation rate is $1\,$Hz. RMS values measured with sampling rate of $10\,$Hz}
\end{table}
  
\subsubsection{Injection box}

\begin{figure}[!htpb]
\includegraphics[width=8cm]{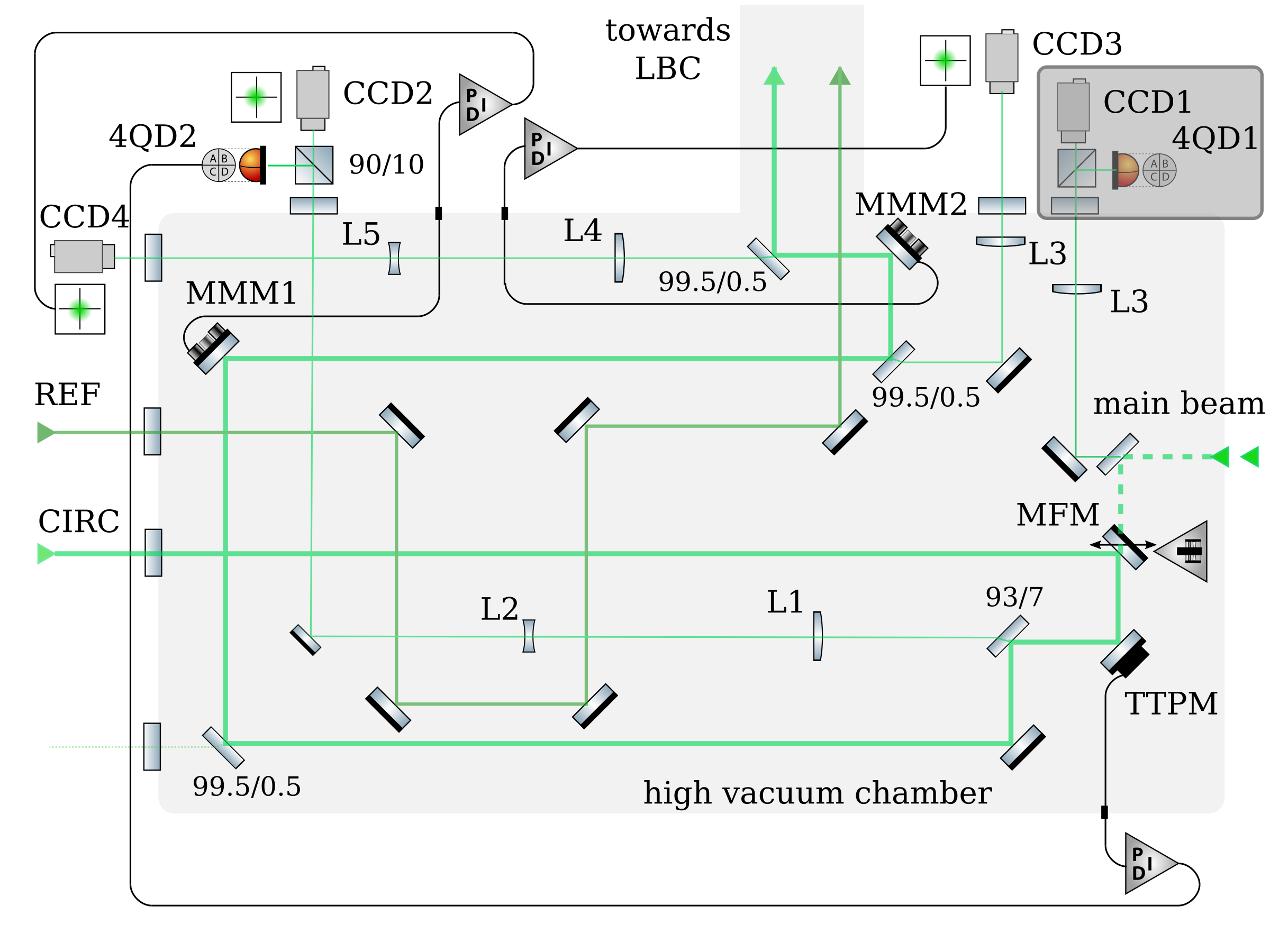}% Here is how to import EPS art
\caption{\label{fig:IB} Optical setup and laser beam path in the Injection Box vacuum chamber used to dispatch the laser beams in the LBC and towards diagnostic tools. MFM: motorized high accuracy flip mirror, TTPM  tip/tilt piezo mirror. MMM1,2: piezo-motorized mirrors. 4QD1,2: 4 quadrant position detectors.  L1, L3, L4:  plano-convex lenses. L2, L5: plano-concave lenses (See text for details).}
\end{figure}

\begin{figure}[!htbp]
\includegraphics[width=8cm]{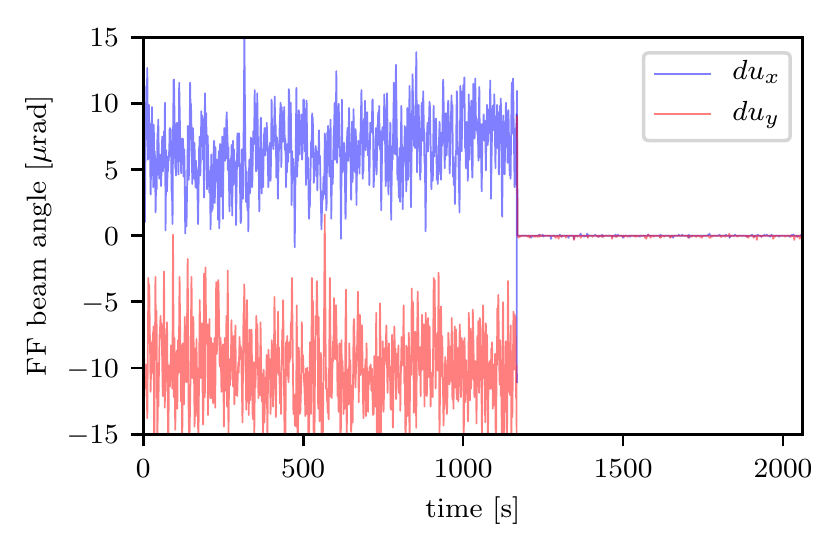}
\caption{\label{fig:stabIB1} Residual pointing instabilities recorded with CCD2 at $10\,$Hz when the fast-pointing stabilization loop is activated on the mount TTPM at $t=1170\,$s}
\end{figure}

The injection in the LBC requires an accurate control of the beam pointing \cite{Dupraz:2014aa,Dupraz:2015ab,Ndiaye:2019aa} of both the main laser beam and CIRC/REF beams. A high vacuum chamber hosts a de-coupled optical breadboard equipped with several leakage mirror pick-ups for the laser-beam diagnostics, a beam dump and beam dispatching in the LBC. A schematic view of the optical setup is given in Fig.~\ref{fig:IB}.
 % figure scheme of the IB
After the injection in the IB vacuum chamber the pointing of the CIRC beam is stabilized by a fast piezo-electric driven $2"$ steering mirror mount (TTPM) made of a 4 piezo-electric stacks giving a high stiffness and high resonant frequencies  $<700\,$Hz able to induce a maximum tilt of $1.5\,$mrad \cite{MRCsystem}. A beam leakage of less than $7\%$ is focalized onto a 4-quadrant photodiode detector (4QD2) with a size of  $(10\times10)\,$mm$^{2}$ thanks to two lenses of focal lengths $f_1 = 400\,$mm and $f_2 = -30\,$mm set at distances to result in an equivalent focal length in excess of $8.9\,$m. The 4-quadrant gap is $90\,\mu$m \cite{MRCsystem}. A closed-loop controller continuously corrects for deviations of the laser beam from the desired position. Fig.~\ref{fig:stabIB1} shows the residual pointing instabilities when the closed-loop controller is activated. By Fourier analyzing the residual noise, it is found that perturbations up to $380\,$Hz are well corrected.  Without activation of this fast-stabilization loop the residual pointing instabilities measured at $10\,$Hz are $du_x= 3.6\,\mu$rad and $du_y=3.7\,\mu$rad.
%figure - IB.FF1 pointing stability

\begin{figure}[!htbp]
\includegraphics[width=8cm]{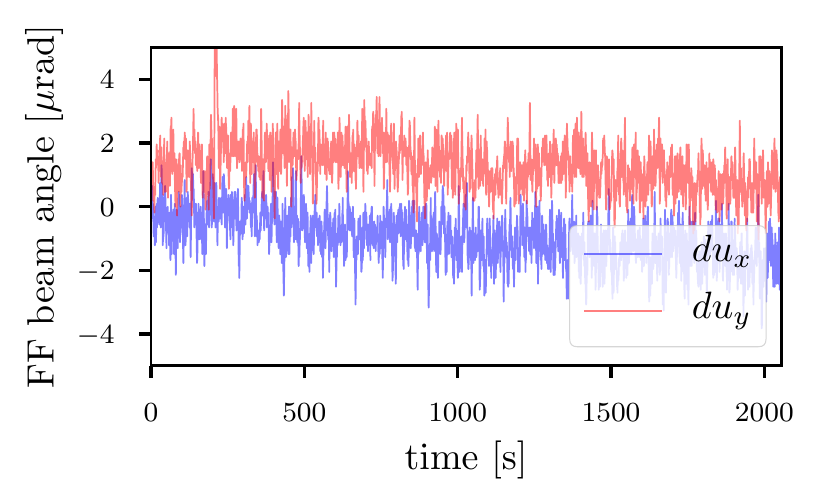}% Here is how to import EPS art
\caption{\label{fig:stabIB2} Residual pointing instability recorded with CCD4 at $10\,$Hz when the laser beam regulation on MMM2 is not activated.}
\end{figure}

The CIRC beam is further reflected by high reflectivity mirrors held in stick-slip piezo-electric motorized mounts MMM1 and MMM2. They control the laser beam injection in the LBC. Its position is controlled on the near field, visualized with CCD3, by imaging the laser in an equivalent plane of MMM2 with a lens of focal length $f_3 = +300\,$mm. The laser beam position is set with an accuracy of $\approx 19\,\mu$m on the near field (L3,CDD3). Its angle is controlled  on the far field, visualized with CCD4, with lenses of focal lengths $f_4=+400\,$mm and $f_5=-50\,$mm giving an equivalent focal length in excess of $7\,$m. The laser beam angular injection is monitored on the far field (L4,L5,CDD4) with an accuracy of $0.8\pm0.1\,\mu$rad. Both steering mirror mounts  are equipped with optical encoders and operated with a closed-loop controller. The resolution of the encoder is $0.17\,\mu$rad. Slight linear drifts of $du_x/dt = - 44\,$nrad/min and $du_y/dt = -31\,$nrad/min are noticeable on Fig.~\ref{fig:stabIB2} when the laser beam regulation is not activated on the far-field. This drift has been identified to be related to a thermal drift on the optical encoder.

%figure - IB.FF2 pointing stability
With all the pointing stabilization loops activated in the SALT and IB systems the residual pointing instabilities measured in air are $dx = 14.1\,\mu$m and $dy =11.9\,\mu$m RMS at the interaction point by the IP imaging optical system. These values fall down to $dx = 2.9\,\mu$m and $dy=3.1\,\mu$m when the system is pumped down and reaches the thermal equilibrium. 

\subsubsection{Interaction point imaging optical system}

\begin{figure}[!htpb]
\includegraphics[width=8cm]{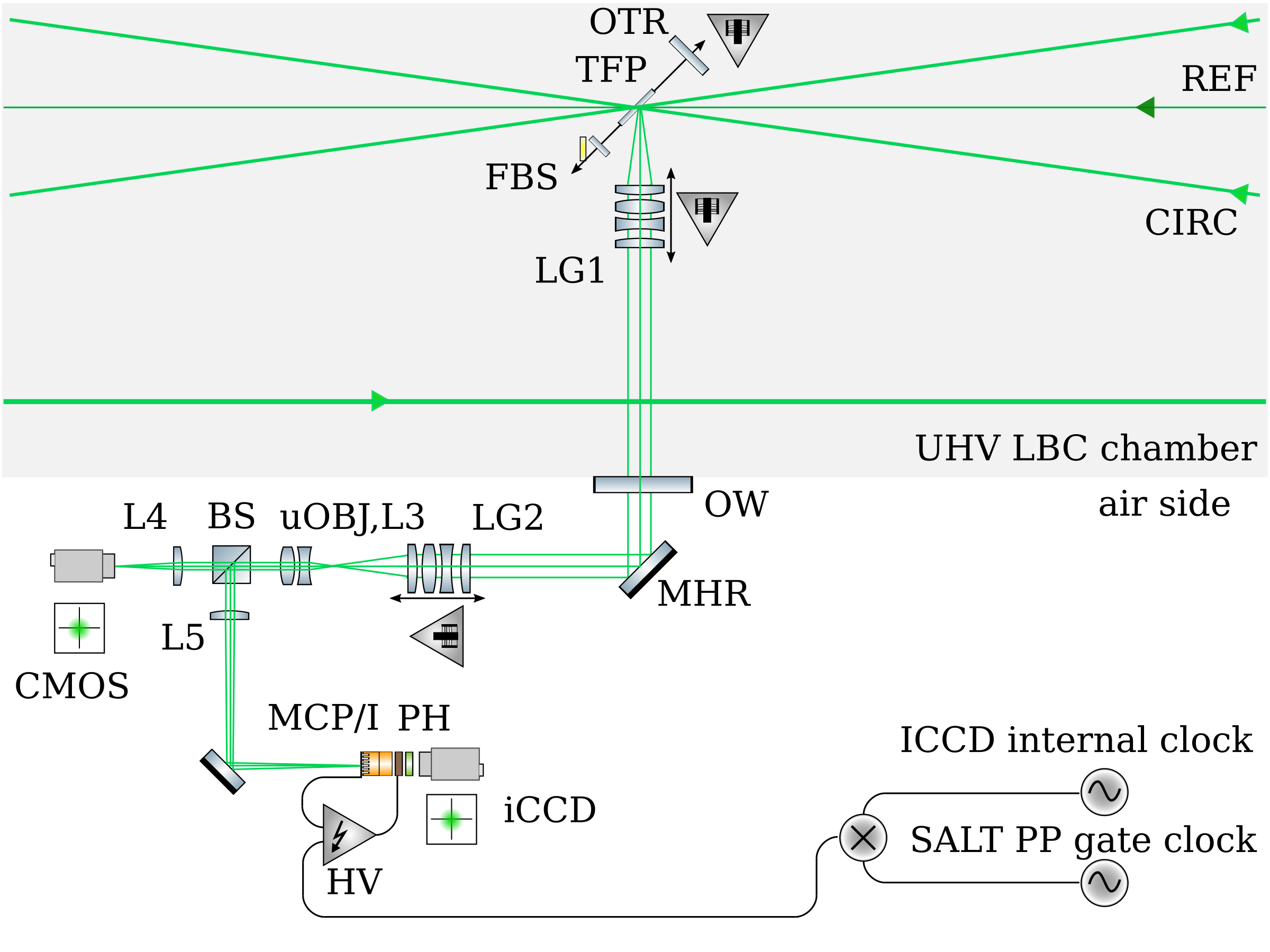}% Here is how to import EPS art
\caption{\label{fig:FAD01} Schematic of the IP imaging optical system. OTR: optical transition radiation target, TFP: thin film plate, FBS: fluorescent beam screen, LG1: Lens group objective 1, OW: optical window, MHR: mirror high reflectivity, LG2: lens group objective, uOBJ,L3 :
high magnification power micro objective, BS: beamsplitter, L4,L5: positive lens, MCP/I: micro channel plate/ intensifier, PH: phosphor, ICCD: intensified CCD, HV: high voltage, CMOS:  Complementary metal oxide semi-conductor area detector.}
\end{figure}

The IP imaging optical system (IIOS) allows to realize the spatial alignment of the LBC passes with a micrometer spatial resolution and temporal resolution of less than $5\,$ns. Thus for each pass the laser-beam profile can be extracted.  It is additionally used to image the interference fringes between REF and CIRC beams at the IP for the synchronization of the LBC.  As shown in Fig.~\ref{fig:FAD01}, a motorized mechanical support holds a thin film pellicle (TFP), reflecting not more than $1\%$ of the incident light depending on the CIRC and REF beam polarization. The reflected beams are used for alignement and synchronization purposes with the CIRC and REF beams. A YAG:Ce fluorescent screen (FBS) and an optical transition target (OTR) made of a fused silica window in order to stand for the large electron beam intensity~\cite{Marongiu:2017aa,Cioeta:2017aa} can be inserted at the IP and  are additionally employed to image the electron beam
A first objective (LG1), placed under vacuum, is fixed on the targets support at the center of the LBC.  A second objective (LG2), located in air, allows to image the IP region through an additional high magnification microscope objective onto two detectors, a CMOS GIGE camera and an intensified gated camera (ICCD).  All the optics sit on opto-mechanical fine-height adjustment supports.

 % figure scheme of the FAD

The IIOS design is driven by spatial and environmental constraints and  the requirement of a high resolution. It relies on an image relay with large numerical aperture objectives (LG1,LG2) to image the IP region outside of the vacuum vessel with a good module transfer function (MTF). In air the image is obtained on the area detectors equipped with standard high resolution microscopes.
High-resolution long focal-length objectives are essential components of quantum gas experiments. There is an increasing need for high resolution imaging, below $3\,\mu$m, in order to extract fine details such as topological defects in Bose-Einstein condensates or single trapped ions \cite{Kawaguchi:2012aa, Streed:2012aa}. The design of the long focal length objectives (LG1, LG2) of the IIOS are mostly inspired by recent works of L. M. Bennie et al.~\cite{Bennie:2013aa}  and J. D. Pritchard~\cite{Pritchard:2016aa}. The design is using that of the Pritchard ATOM objectives modified to provide at a central wavelength of $515\,$nm an infinity-corrected and diffraction limited performance whilst maximizing the available numerical aperture given the minimum working distance of $60\,$mm imposed by the space inside the LBC corona and the TFP inserter. For reasons of simplicity, all elements are constrained to be of $2\,$inches diameter BK7 singlets. The objectives have been iteratively designed using the ray-tracing software Zemax\texttrademark\ \cite{Zemax:software} to numerically optimize the lens spacing while minimizing the RMS wavefront error at the wavelength $\lambda_0$. 

 % figures spot of the FAD
\begin{figure}[!htpb]
\includegraphics[width=8cm]{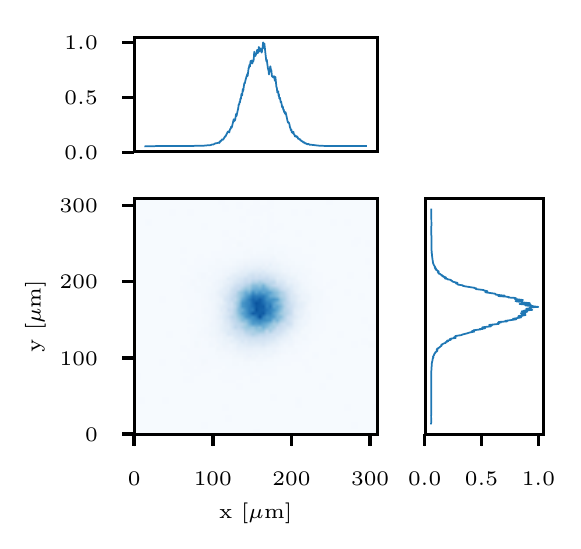}\\% Here is how to import EPS art
\includegraphics[width=8cm]{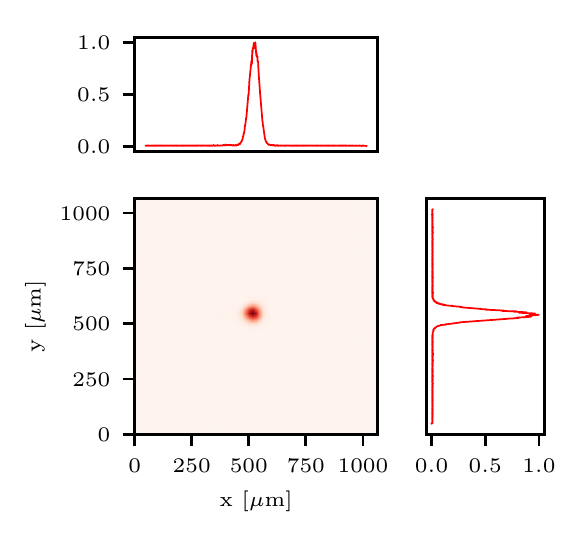}
\caption{\label{fig:FAD-spot} (Top) image of the laser spot of the first pass at the IP taken with the ICCD with a gain of 50, $t_{gate} = 2.7\,$ns, 500 images are accumulation on the CCD. (Bottom) similar image taken with the CMOS area detector with gain of 1 and $20\,\mu$s acquisition duration.}
\end{figure}
 It results that each objective lens has a point spread function with $\sigma \approx 1.1\,\mu$m.
 The drawback of using this image relay is the inherent sensitivity to alignment especially since the first objective $LG1$ stands inside the vacuum chamber and the second objective $LG2$ after the vacuum optical window. In the present design a magnification of $M_{1,2} = 1$ is chosen due to space constraints. The design of the relay imaging outside the vacuum chamber allows for more flexibility and the possibility to use a standard microscope configuration. The imaging line magnification is design to achieve to have a resolution close to $1\,\mu$m in the object plan and also to achieve a good resolving power to be able to measure the contrast of the interference fringes for the synchronization of the LBC on $\nu^{(46)}$, the magnification of the microscope objectives has to be large enough to avoid their blurring, due to the limited MTF of the ICCD. Effectively, the combination of micro-channel plates (MCP/I), spaced by less than $6\,\mu$m, and the phosphor grain (PH) considerably degrades the MTF of the intensified time-gated detectors. One of the best combination is used and a $MTF_{ICCD}\approx 45\,$lp/mm  \footnote{lp$/$mm is line pair per mm} has been specified by the manufacturer and confirmed during dedicated test.

The period of the interference fringes is $i=\lambda_0/\sin\phi= 3.7\,\mu$m.  The minimum microscope objective power (uOBJL3) is then driven in first approximation by the sensor resolution and a sufficient number of points per fringes giving $P_{obj} ^{ICCD} \sim 30 $ for the ICCD imaging line.  The same argumentation considering a pseudo MTF$_{CMOS} \approx 91\,$lp/mm for the CMOS detectors with pixel of $5.5\times5.5\,\mu$m$^2$ leads to $M_{obj}^{CMOS} \sim 11$. A  $\times50$ microscope objective (uOBJL3) with a front focal length $f_3 = +4\,$mm coupled to a plano convex singlet with back focal length of $f_4= 40\,$mm and $f_5 = 125\,$m have been chosen for the CMOS and ICCD detectors respectively. Fig.~\ref{fig:FAD-spot} shows the images of the laser spot of the first pass at the IP. Table~\ref{tab:IIOS} summarizes the IIOS parameters.

 \begin{table}[!htpb]
\caption{\label{tab:IIOS} IP imaging optical system parameter}
\begin{ruledtabular}
\begin{tabular}{llcc}
parameters &  ICCD det.  & CMOS det.  & unit \\
\hline
field of view	& $0.3$ 	 &	 $1$	& mm \\
resolution  	& $0.30$ &  	$0.54$	& $\mu$m/pixel \\
acquisition time          &  $2.72$ 		 &  	$2400$	&	ns \footnotemark[1]\\
spectral bandwidth 	&  $\pm 1$		& 	 $\pm 1$		& nm  \footnotemark[2] \\
%noise level		&  $10\pm5$		& 	 $1\pm1$		& cc/pixel  \\
\end{tabular}
\end{ruledtabular}
\footnotetext[1]{Minimum acquisition time, the maximum fps are $4\,$Hz and $10\,$Hz respectively}
\footnotetext[2]{Given by the optical design and on which the resolution is achieved}
%\footnotetext[3]{given by the optical design and on which the resolution is achieved}
\end{table}

\section{\label{sec:TestConditions} Optical commissioning conditions}

The tests have been realized in the assembly hall, of ISO8 cleanliness, of the ALSYOM/SEIV-ALCEN company in collaboration with LAL/CNRS. The IP module has been assembled under an ISO5 cleanliness mobile air ceiling. The cleanliness requirement is driven by the future high power laser beam exposure of optical elements of the LBC and the IB. The volume of the ISO5 air ceiling is $(4.5 \times 4 \times 5)\,$m$^3$. Its flow is $32 000\,$m$^3$/h at a speed of  $0.45\,$m/s. The ambient noise generated by the air ceiling is in excess of $69\,$dB. The hall also hosts at least five other ISO5 air ceilings of various sizes, two dust cleaning showers and a stand for UV dust inspection. All these equipments generate a significant acoustic noise. This environment is thought to be equivalent or more noisy than that of the accelerator bay where the IP module will finally be installed and where chillers used for the cooling of accelerating structures will be running continuously.  
The ISO8 hall is air-conditioned with a temperature and relative humidity regulations. The setting temperature is $T_0\approx 20\,^\circ$C and can be varied in the range from $19^\circ$C to $22^\circ$. The temperature has been monitored in various points of the IP module. The temperature under the air ceiling $T_{LBC}$ was oscillating in the range [$20.5^\circ$C,$22.0^\circ$C] over the test period. Under the air ceiling an optical table has been installed for the MPS setting and measurement. To reduce the effect of air turbulence during the laser-beam angular deviations measurement with the autocollimator, a removable enclosure has been installed over the optical table. The temperature on the optical table in the enclosure was $T_{MPS} = T_{LBC}+0.5^\circ$.

The aim of the commissioning is the validation of the installation process, the validation of the pre-alignment strategy, the implementation of alignment and synchronization algorithms and the demonstration of the performances with a low power laser beam simulating the electron and the high power laser beams. 

\section{\label{sec:LBCprealignment} Pre-alignment of the laser beam circulator}

The LBC pre-alignment entails the definition of its optical axis, the IP reference point in the mechanical reference network defined by the alignment of the M1 parabolic mirror, the alignment of M2 in confocal arrangement with respect to M1, the definition of the reference position of the IP on the IIOS and the MPS alignment and installation in the LBC. 

\subsection{Optical axis and parabolas alignment}

The tolerance on the angle of the LBC with respect to the electron axis is $1.7\,$mrad, on the basis fixed by the required performances on the $\gamma$-ray flux and bandwidth (TASD),  and limited by the positioning of the optical elements and knowledge of the positioning of the optics on its mounts. When the IP module will be integrated in the GBS low-energy linac, the residual angle error between the electron beam and the LBC optical axis will be the sum of the cavity beam position monitors positioning error in the mechanical reference network and the error of the LBC optical axis orientation

The optical axis of the LBC is defined by the collimated REF beam that is sent through a $6\,$mm-diameter pinhole (P1) fixed on the mount base of M1 (see Fig. \ref{fig:LBCblocks}), which generates a diffracted beam, to a $1\,$mm-diameter pinhole (P2) fixed on the mount of M2. On the backside of P2 a near-field laser-beam profiler imaging system (P2-NF) and a far-field beam pointing diagnostic (P2-FF) are placed. P1 and P2 are positioned with the help of a laser tracker with a precision of $\sigma_{LT}\pm 20\,\mu$m.  Once the two pinholes are positioned with the laser tracker in the reference mechanical frame, the REF beam is aligned and reference marks are taken in the (P2-NF, P2-FF).  The precision of the centering of the diffracted beam on P2-NF is $\sigma_{C} = \pm 35\,\mu$m which is mainly limited by the ambient noise and resolution on the P2-NF. A couple of near-field and far-field diagnostics (P1-NF, P1-FF) is further added in between the pinhole P1 and the mechanical position of the IP. 
Once the axis is defined, the numerical reference marks for the IP coordinates are taken on the IIOS cameras.

\begin{figure}[!htpb]
\includegraphics[width=8cm]{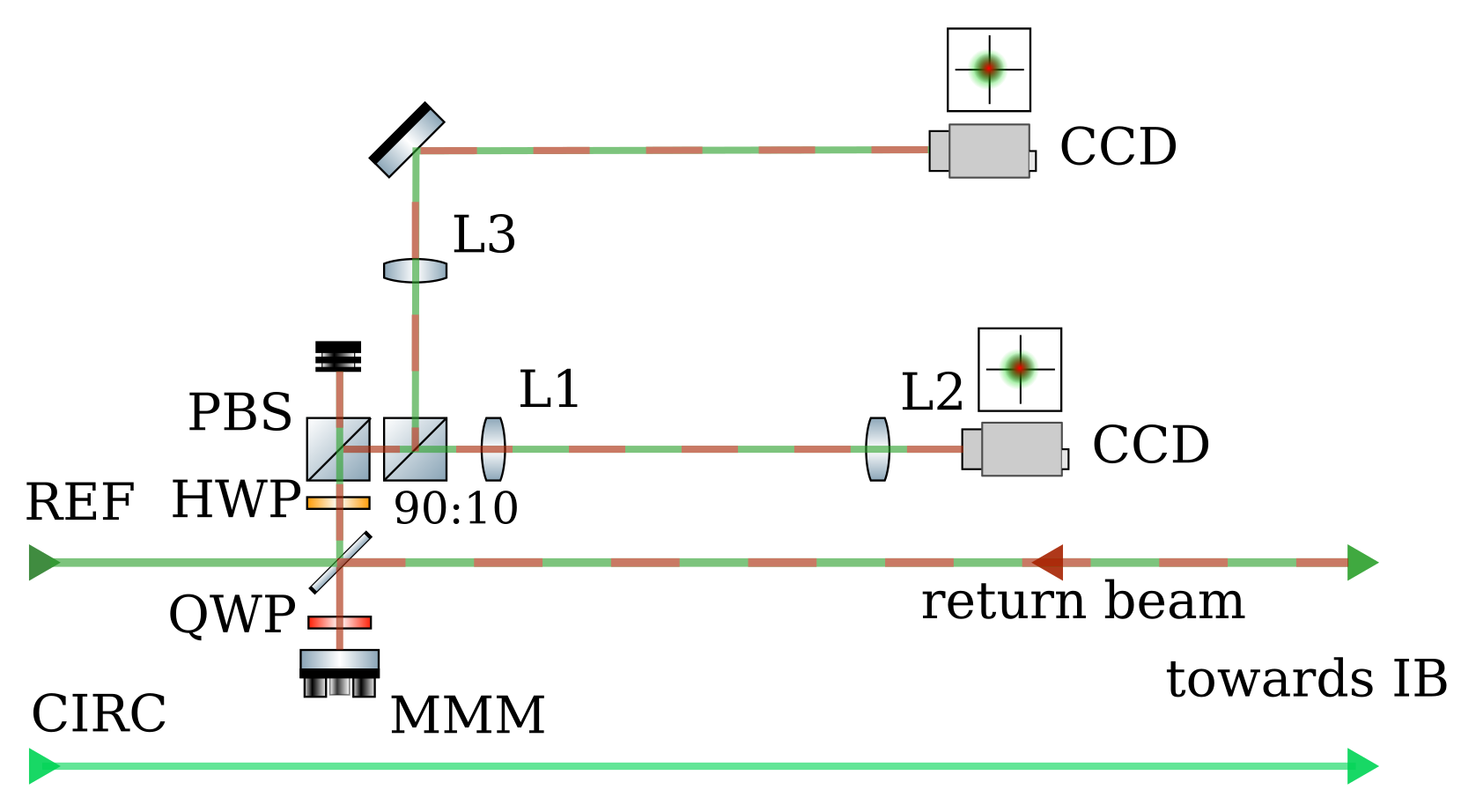}
\caption{\label{fig:SRA}  Optical scheme of the self referenced autocollimator. HWP: half wave plate, PBS: polarized beam splitter, QWP: quarter wave plate, MMM: piezo motorized mirror mount, L1,L3: plano-convex lens, L2: plano-concave lens, CCD : area detectors}
\end{figure}

The next step consist in the positioning of M1 optical axis onto the REF axis. 
The  (P2-NF, P2-FF) beam diagnostics are displaced in between the IP and M2. The REF beam axis is then monitored by on two points along in the LBC. A flat $(400\times400)\,$mm$^2$ square mirror is installed on the M2 motorized mount and set in autocollimation on the REF beam. At this point the pinhole  are taken out. 
The autocollimation is measured thanks the self referenced autocollimator (SRA) installed between the IB and SALT instruments in the IP module. It makes a reference out of any incoming beam, and diagnoses any misalignment between the incoming beam and the return beam.  The principle of SRA is illustrated in fig.~\ref{fig:SRA}.
%FIG SRA

Once the flat mirror is aligned in autocollimation on the REF axis, a $50\,$mm-diameter, $100\,$mm focal length spherical mirror mounted on a piezo-electric motor is installed to create a source point at the IP. For this purpose the TFP is turned by $90^\circ$ to collect the REF beam reflected by the spherical mirror in the IIOS. The size of the focal spot generated by the spherical mirror is optimized and it is positioned on the IP reference coordinates previously taken using the IIOS. The TFP is then turned back to its original position. A collimated beam is created by M1 and reflected back, by the flat square mirror and focuses again at the IP. Wherefore, the M1 is aligned on the defined optical LBC axis and the focusing is optimized by looking to the focal spot positions on the IIOS. At this point the M1 is fixed and becomes the reference optical element of the LBC setup. 

The CIRC beam is then injected and steered to superimpose the first pass, dubbed $p=0$ on the IP reference coordinates of the IIOS. The CIRC beam collimation is slightly adjusted to get the best focusing by using the beam expander (BEX) in the SALT. 
M2 is then installed in its 5-axis motorized mount. The transverse positioning is done in the mechanical network reference thanks to the laser tracker and using the motorized translation stages $TX$, $TY$ and $TZ$ (see Fig. \ref{fig:lbc-5ddl}). The tilts of M2 are set by superposing the second pass, $p=1$, on the reference marks of the first pass in the IIOS. At this point, M2 is aligned in autocollimation but not yet in confocal geometry with M1 as the error along the $z$-axis is still larger than $100\,\mu$m. The error budget for the pre-alignment is summarized in the Table~\ref{tab:AlignBudgetError}. 

%TAB PRE ALIGNMENT ERROR BUDGET
 \begin{table*}[!htpb]
\caption{\label{tab:AlignBudgetError} Alignment budget error for elements of the LBC and accuracy of methods and tools used.  All $\mu$m and $\mu$rad}
\begin{ruledtabular}
\begin{tabular}{llcccl}
elements & axis & tolerance errors & total errors  &  accuracy & used method$/$tools \\
\hline
pinhole P1 & $x,y,z$ & $\pm50$ & $\pm28$ & $\pm20$ & LT \\
pinhole P2 & $x,y,z$ & $\pm50$ &  & $\pm 38$ & P2-NF + P \\
beam centering PX-NF & $(x_1,y_1);(x_2,y_2) $ & ($\pm35$,$\pm35$) & ($\pm37$,$\pm37$)  & ($\pm35$,$\pm35$)  & (P1-NF, P2-NF) \footnotemark[1] \\
beam pointing PX-FF & $(u_1,v_2); (u_2,v_2)$ &  ($\pm30$,$\pm30$) &  ($\pm21$,$\pm21$)  & ($\pm18$,$\pm18$)  & (P1-FF, P2-FF) \footnotemark[2]  \\
 \hline
M1  $+$ mount & $x,y$ & $\pm500$ & $\pm450$ &  $\pm450$ & mounting, LT \footnotemark[3]\ \\
M1 tip/tilt & $\theta_x$, $\theta_y$ & $<1700$ & $\pm18$ & $\pm10$ & auto collimation (SRA) \\
M1 focus & $z$ & $<100$ & $\pm 17$ & $\pm 14$ & focusing full aperture \\
M2 $+$ mount  & $x,y$ & $\pm500$ & $\pm 450$ & $\pm 450$ & mounting, LT \\
M2  focus & $z$ & $ - $  & $\approx 3000$ & $2740$  & Focusing Rayleigh length scan  \\
M1-M2 autocollimation & $\theta_x$, $\theta_y$ & $<100$ & $\pm14$ & $\pm3$ & first pass circulation \footnotemark[4] \\
M1-M2 confocality & $z$ & $<100$  & $<100$ & $\approx 90$  & first pass synchronization  \\
 \hline
$\delta D_{MPS}$ &  & $<100$ & $+140/+120$  & calibrated block\footnotemark[5]  \\
$\delta \theta_0$ & & $\pm 3400 $ & $\pm 1700$ & indexing \footnotemark[6]  \\

\end{tabular}
\end{ruledtabular}
\footnotetext[1]{Error and accuracy determined by the image analysis to determine the center of the diffraction pattern and the centroid of the spot.}
\footnotetext[2]{Angular error limited by the beam pointing stability due to air turbulence and accuracy limited by the short focal length used}
\footnotetext[3]{The precision on the positioning of M1 and M2 is limited by mount. The budget is fixed by the paraxial approximation and aberration produced by the tilt to compensate for the decentering of M1 and M2.}
\footnotetext[4]{The final budget is limited by the beam pointing stability in air. }
\footnotetext[5]{The precision on the indexation MPS angle $\theta_0$, the MPS mirrors distance error $\delta D_{MPS}$, the precision of the synchronization $c\delta t_{\rm sync}$ and the tuning of the laser frequency contribute quadratically to the total uncertainty.}
\footnotetext[6]{A systematic shift of $\sim130\,\mu$m has been measured due to the calibrated block definition error used to set this length. It was corrected after the MPS installation by applying a systematic rotation of $\delta \theta = 3.97\,$mrad of each MPS}
\end{table*} 

To set M2 in confocal position with respect to M1 focus point, the first MPS, MPS1, is installed in the LBC. The frequency of the laser oscillator is adjusted to correspond to the frequency $\nu_{air}^{(46)}=\nu_{rf}/(46\times n)$ of an air circulation round trip that would match the theoretical round trip duration in vacuum. At the zero position the MPS added path length is $l = D_{MPS} \cos{(\theta_0)}$. The remaining error comes from $D_{MPS} $ which is $\sigma_{D_{MPS}}  = 0.1\,$mm, then $\sigma_l = \sigma_{D_{MPS}} \cos{(\theta_0)} \approx 91\,\mu$m. The synchronization is scanned for $p=1$ by moving the $TZ$ axis of the M2 motorized mount. The final position in $z$ of M2 has been monitored by the laser tracker and a shift of $z=+2.1\,$mm is measured at the end of the procedure. This discrepancy may be explained by the poor accuracy of the knowledge of the vertex coordinates of the parabolic mirrors in their mounts, see Table~\ref{tab:table2}. 

\subsection{Mirror pairs system parallelism setting}
\begin{figure}[!htpb]
\includegraphics[width=8cm]{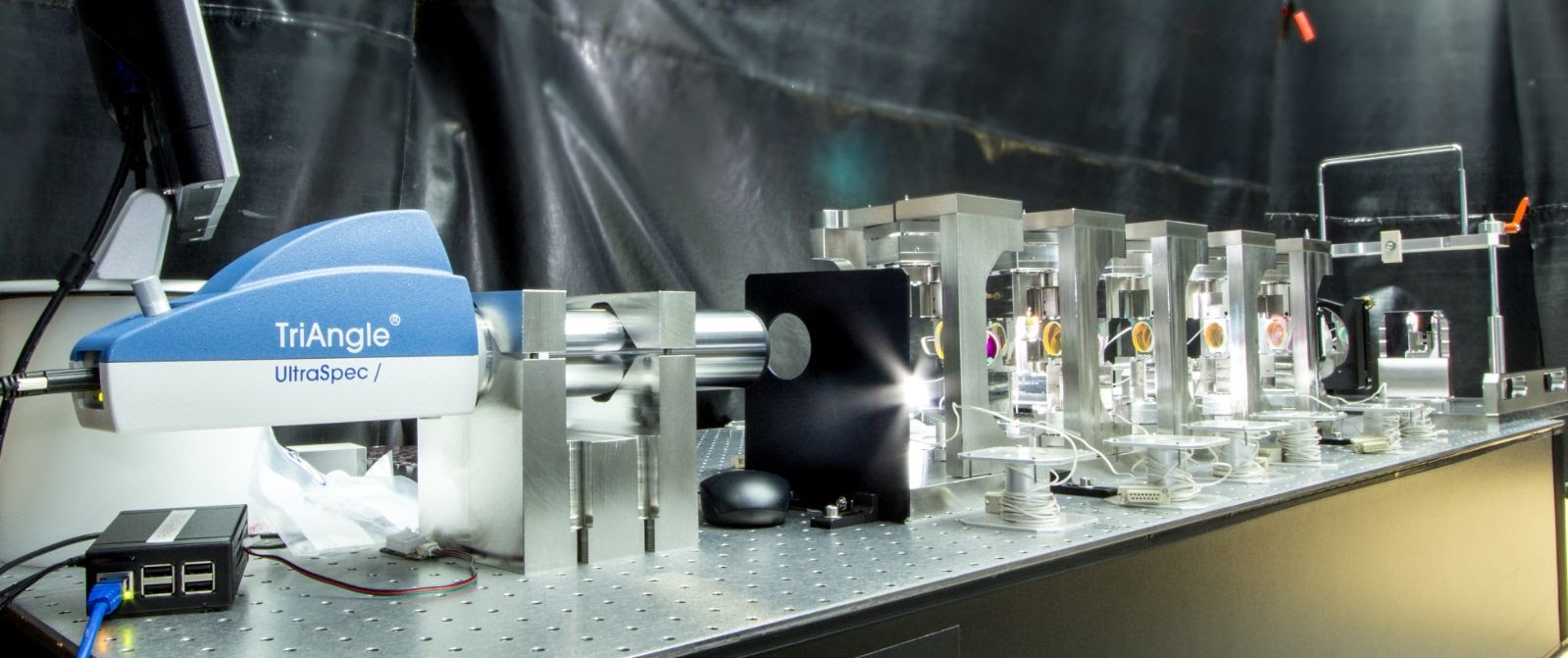}
\caption{\label{fig:MPS-photocascade}  View of the cascade test bench for the MPS before the installation of the top cover. On the left the autocollimator,  in the center the MPS mounted alternatively on the optical and on the upside down support and on the right the reference flat mirror.}
\end{figure}

After the glueing process, the MPS are mounted on their piezo-electric rotation stage and base. This ensemble is installed in the test bench where the autocollimator is set (Fig.~\ref{fig:MPS-photocascade}). The rotation stage indexing at $\theta_0$ is done by the locating base and by observing the angular horizontal deviation of the back side of the MPS mirror thanks to the autocollimator. The yaw of the base is then set by looking at the vertical angular deviation in the autocollimator. Each MPS parallelism is tuned up within the tolerance and stored under a temperature controlled enclosure. 

%FIG MPS CASCADE 
\begin{figure}[!htpb]
\includegraphics[width=8cm]{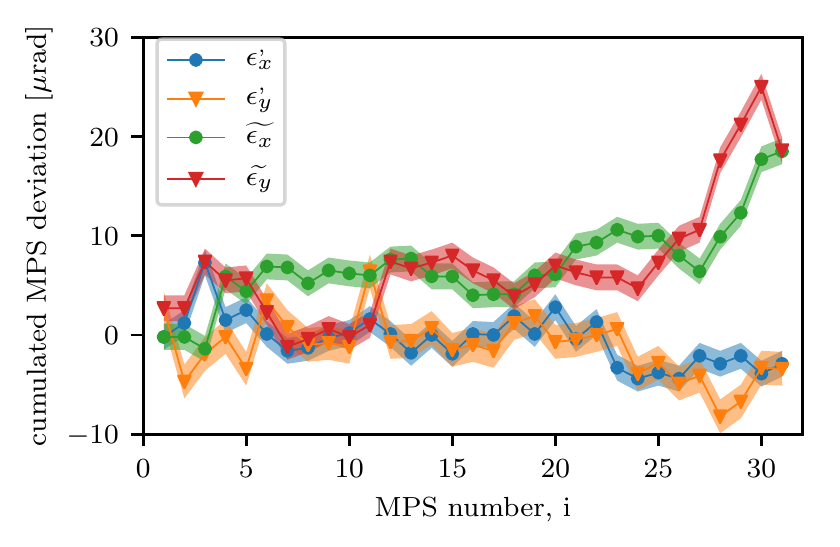}
\caption{\label{fig:MPS-cascade}  Measured cumulated angular beam deviation with ($\epsilon^,_x, \epsilon^,_y$) and without ($\widetilde{\epsilon}^,_x,\widetilde{\epsilon}^,_y$) the cascade cumulated deviation error correction.}
\end{figure}
The MPS are then installed in cascade in the bench. For space constraints in the bench a maximum of 11 MPS can be installed. Each time a MPS is added in the cascade bench, the angles of the newly inserted MPS are adjusted to keep the total deviation $\epsilon' = \sum_i \epsilon_i$ within the tolerance, see Table~\ref{tab:paramIP}. 
%FIG MPS CASCADE 

The results are shown in Fig.~\ref{fig:MPS-cascade}.  The total cumulated angular deviation is $\epsilon' = 4.5\pm1.0\,\mu$rad with cascade correction and $\widetilde{\epsilon}' = 26.2\pm1.0\,\mu$rad without. As the autocollimator light goes forth and back in the MPS, the parallelism error is the half of the measured angular beam deviation. The MPS parallelism setting has been done at a temperature of $T_{MPS} = 21.2\pm0.1^\circ$C.
 
\section{\label{sec:LBCresults}Laser-beam circulation results}

\subsection{Laser beam circulation and mirror pair system installation}
\begin{figure}[!htpb]
\includegraphics[width=8cm]{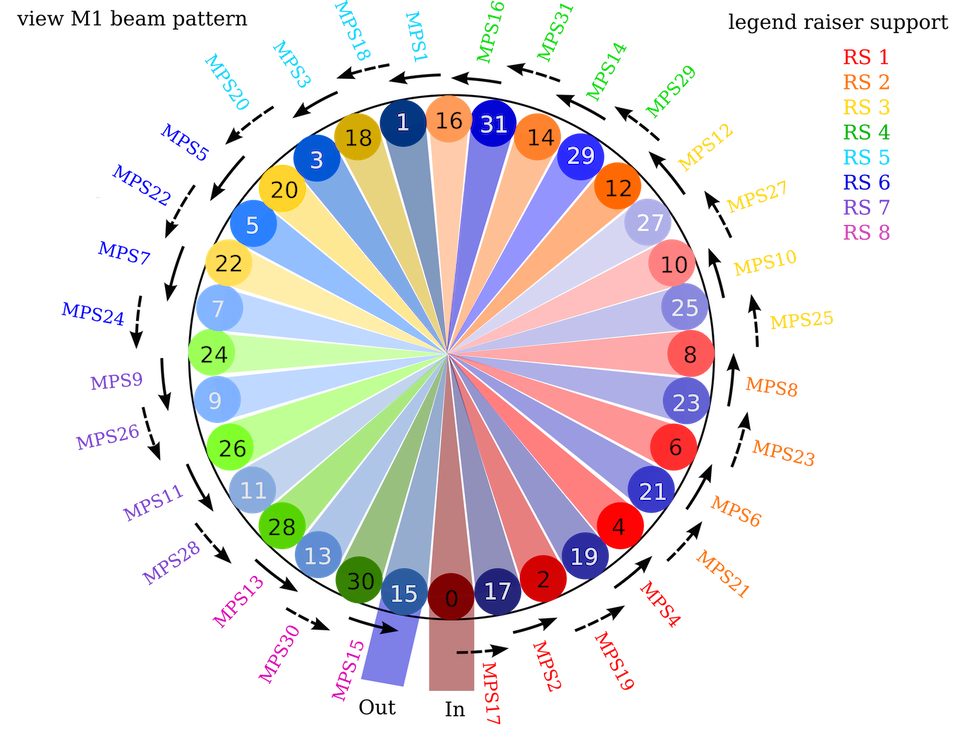}\\% Here is how to import EPS art
\caption{\label{fig:LBC-CIRC-SCHEME} Schematic view of the laser beam circulation on the M1 parabolas seen from the IP point. The \textbf{RS \textit{n}} are the support of the MPS and injection/ejection mirror pairs,  indexed in colors, and the arrow represent the MPS. 4 MPS are mounted on each raiser support except for the first and the last ones where the injection and ejection mirrors are fixed respectively}
\end{figure}

The laser beam impact does a $4\pi$ rotation on the parabolic mirrors as shown on the Fig.~\ref{fig:LBC-CIRC-SCHEME}. The beam impinges M1 thus making the first impact $\#0$ on the parabola, the reflected beam is focused at the IP and is collimated back by M2 towards M1 through the MPS1 and generates the impact $\#1$ on the M1 parabola. The processus is then repeated 31 times.

%FIGURE CIRCULATION

\begin{figure*}[!htpb]
\includegraphics[width=16cm]{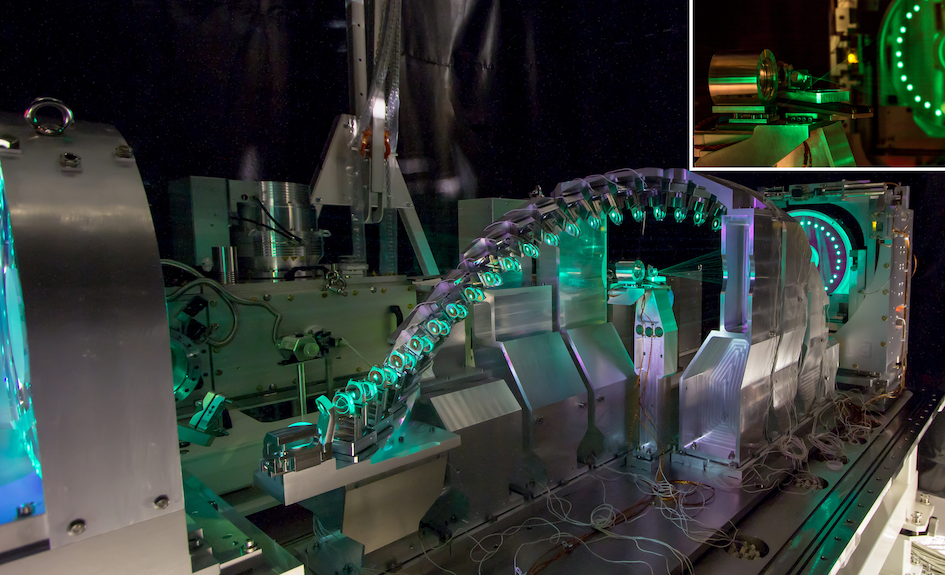}\\% Here is how to import EPS art
\caption{\label{fig:LBC} Pictures of the LBC in air with all the MPS installed before the fine tuning of the alignment. M1 is placed on the left and M2 on the right hand side of the picture. At the center the TFP is retracted and the first UHV objective can be seen. A magnification of this IP region is given in the inset. Courtesy of N. Beaugerad, SEIV.}
\end{figure*}

The MPS$i$ have been installed in the optical order with $i=\llbracket1,31\rrbracket$.  Note that the indexing $i$ corresponds to the optical order and equals the pass number $p$. The mechanical index of thx e MPS is different on the helix made by the raiser support as one can see on the figures Fig.~\ref{fig:LBC-CIRC-SCHEME} and Fig.~\ref{fig:LBC}. One can notice that the MPS16 is in the middle of the LBC. The advantage of the installation in the mechanical order is that each time a MPS is installed there is free space on one of its side. The main disadvantage is that $17$ MPS must be installed in order to achieve a first circulation of the laser beam at the IP.  Given the mechanical sensitivity of the MPS to shocks, the optical order installation has been chosen so that each time a MPS is installed a new pass is generated at the IP. An online monitoring of the spot position in the IIOS is performed during installation and thus any shock on a newly installed MPS can be immediately detected. The MPS installation is a very delicate handling task especially due to the very limited free space after the MPS16 has been installed.  

%PHOTO ALL MPS INSTALLED 

The successive centroids $(x_{c}^{(p)}, y_c^{(p)})$ of the laser spots recorded at each pass $p$, is dubbed circulation trace. During the installation of the first 16 MPS a very good superposition of all the passes, i.e. a flat circulation trace, has been observed after a quick alignment optimization of the injection and the M2 rotation axes and focus. During the installation of the next 15 MPS, a total of 8 MPS have been unfortunately touched or shocked inducing a significant misalignment ($\epsilon > 15\,\mu$rad) due to the delicate installation operation in the tight space between successive MPS bodies. The flexure blade is indeed extremely sensitive to a mechanical shock on the differential screw which is protruding on the back side of the blade. Consequently every faulty MPS is removed from the support and reset on the autocollimator bench as close as possible to its cascade value setting, for both axes. This required procedure unfortunately multiplies the number of handlings and consequently the risk of shocks and misalignments of other MPS. In addition, a temperature increase in the LBC ISO5 clean area due to human activities constrains the MPS installation time. The parallelism cannot be preserved during the mechanical installation for all the MPS within the tolerances. The maximum misalignment observed in the installed MPS is $\approx 50\,\mu$rad. The average misalignment of the MPS is below $20\,\mu$rad. Consequently some of them must be steered "\textit{in -situ}", using the IIOS as a high resolution autocollimator with the IP coordinates as reference mark. This method has been chosen to finalize the alignement. 

\subsection{Misalignment sensitivity}
\begin{figure}[!htpb]
\includegraphics[width=8cm]{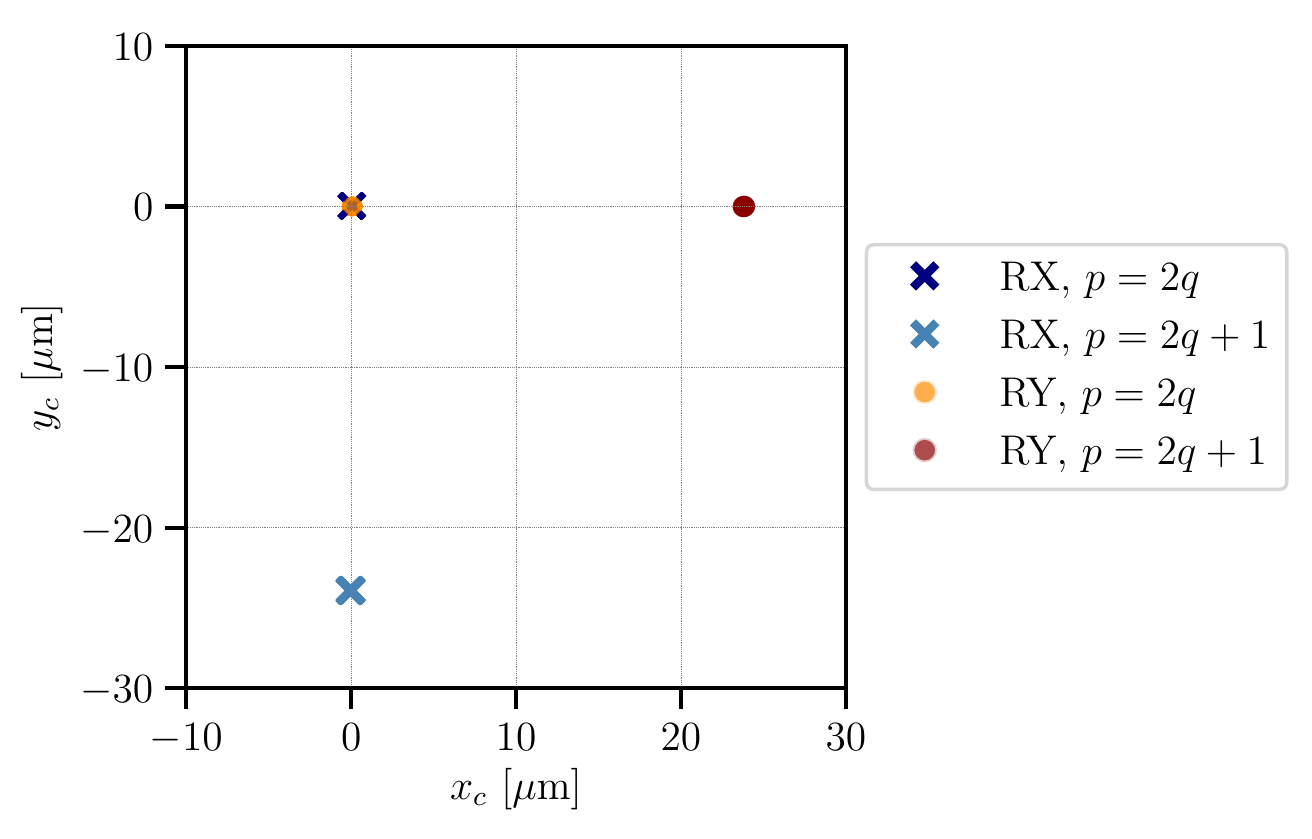}% Here is how to import EPS art
\caption{\label{fig:CIRC-M2-RXRY} Simulation of the effect of a $5\,\mu$rad solely $RX$ or $RY$ rotation misalignment of M2 on the circulation trace at the IP.}
\end{figure}

\begin{figure}[!htpb]
\includegraphics[width=8cm]{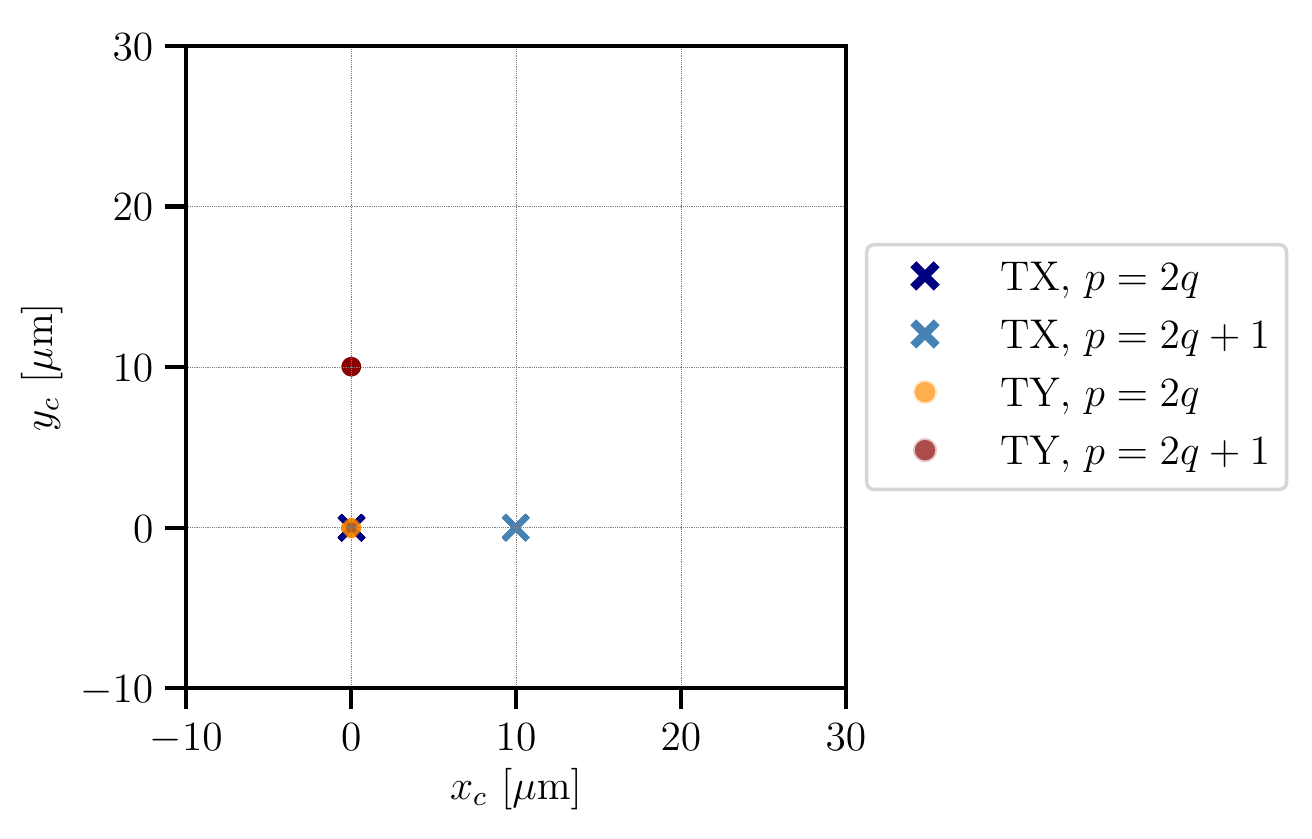}% Here is how to import EPS art
\caption{\label{fig:CIRC-M2-TXTY} Simulation of the effect of a $10\,\mu$m solely $TX$ or $TY$ translation misalignment of M2 on the circulation trace at the IP.}
\end{figure}

\begin{figure}[!htpb]
\includegraphics[width=8cm]{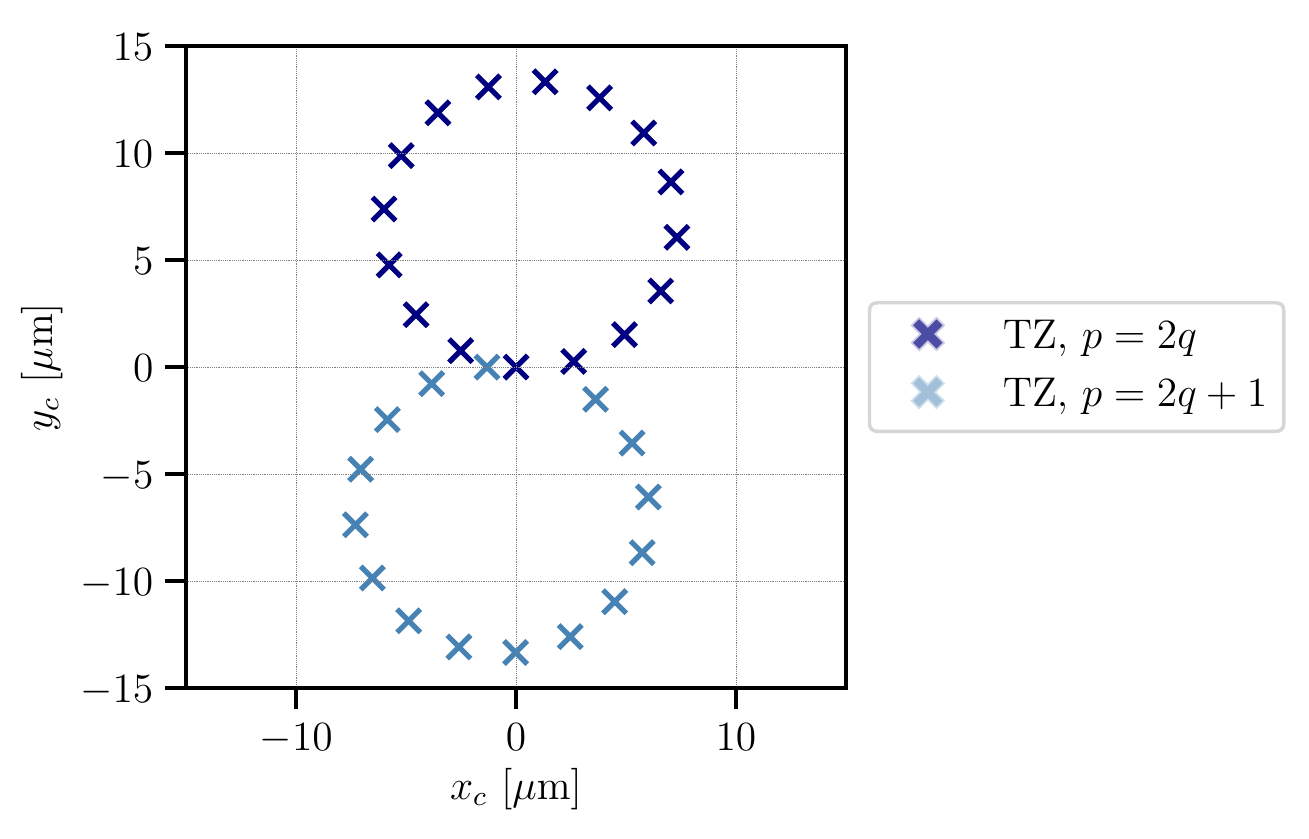}\\%\\% Here is how to import EPS art
\includegraphics[width=8cm]{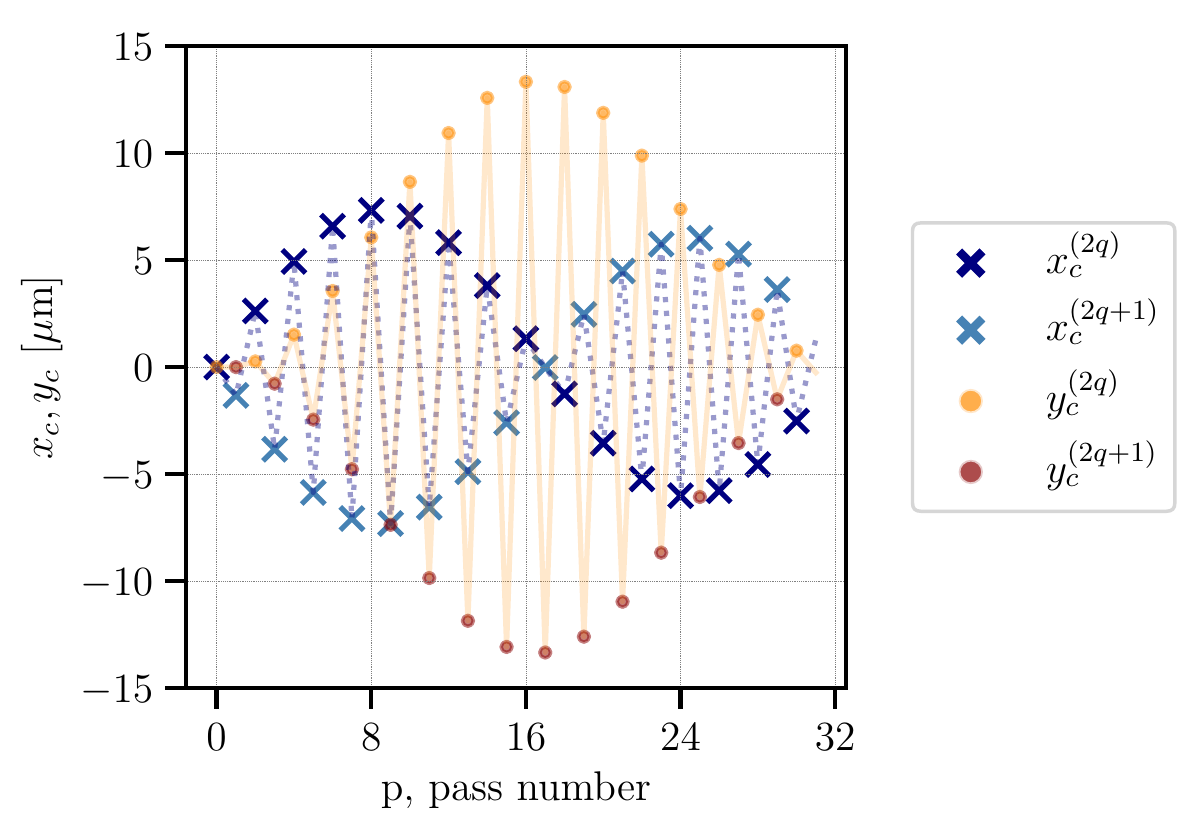}% Here is how to import EPS art
\caption{\label{fig:CIRC-M2-TZ} Simulation of the effect of a $10\,\mu$m solely $TZ$ translation misalignment of M2 on the circulation trace at the IP}
\end{figure}

To give a better understanding of the LBC alignment behavior, the dependence of the circulation trace recorded by the IIOS on the various degrees of freedom -- the injection angle $u_x$, $u_y$, the axes of M2, namely $TX,TY, TZ, RX$ and  $RY$, and the $31$ pair of MPS angles ($\epsilon_x,\epsilon_y$) -- is computed with a model of the system. The transverse translations $TX$ and $TY$ and rotations $RX$, $RY$ are degenerated, as can be deduced from Figs~\ref{fig:CIRC-M2-RXRY} and~\ref{fig:CIRC-M2-TXTY}. 
% FIGURE MISALIGNMENT M2.RX, M2.RY
We consider that the accuracy in the process of the optics manufacturing, positioning in its mount and alignment allow us to make the assumption that the $TX$ and $TY$ are determined and frozen by the pre-alignment accuracy. A compensation for a residual transverse misalignment of M2 by a rotation of M2 does not affect the optical performances as far as the misalignment stays within the paraxial approximation with an off centering error of M1 and M2 $< 0.5\,$mm, which must be the case given the pre-alignment tolerance.
% FIGURE MISALIGNMENT M2.TX, M2.TY
The M2 rotation $RX$ creates two populations in the circulation trace as shown in figure Fig.~\ref{fig:CIRC-M2-RXRY}. The odd laser spot population is shifted by $\sim 2\times ROC_2\times RX$ ($\sim 2\times ROC_2\times RY$) at the IP on the $y$-axis ($x$-axis) while the even laser spots remain unchanged.
A translation $TZ$  along the $z$-axis of the LBC creates a figure-8 at the IP centered on the first pass $p=0$. The odd (even) spots rotate anti clockwise on the top (bottom) part of the trace, as illustrated on~Fig.~\ref{fig:CIRC-M2-TZ}.

% FIGURE MISALIGNMENT TZ
\begin{figure}[!htpb]
\includegraphics[width=8cm]{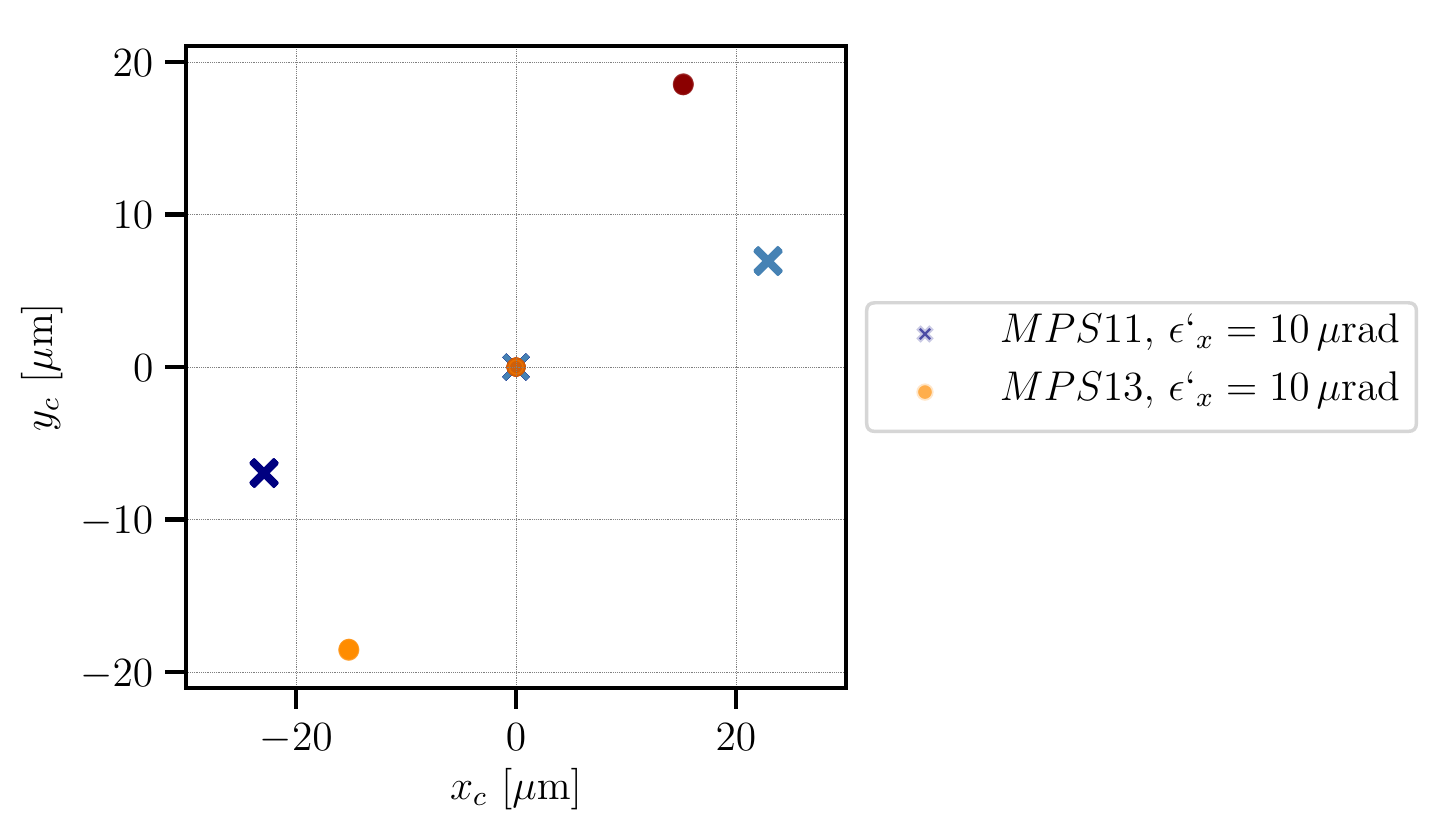}\\% Here is how to import EPS art
\includegraphics[width=8cm]{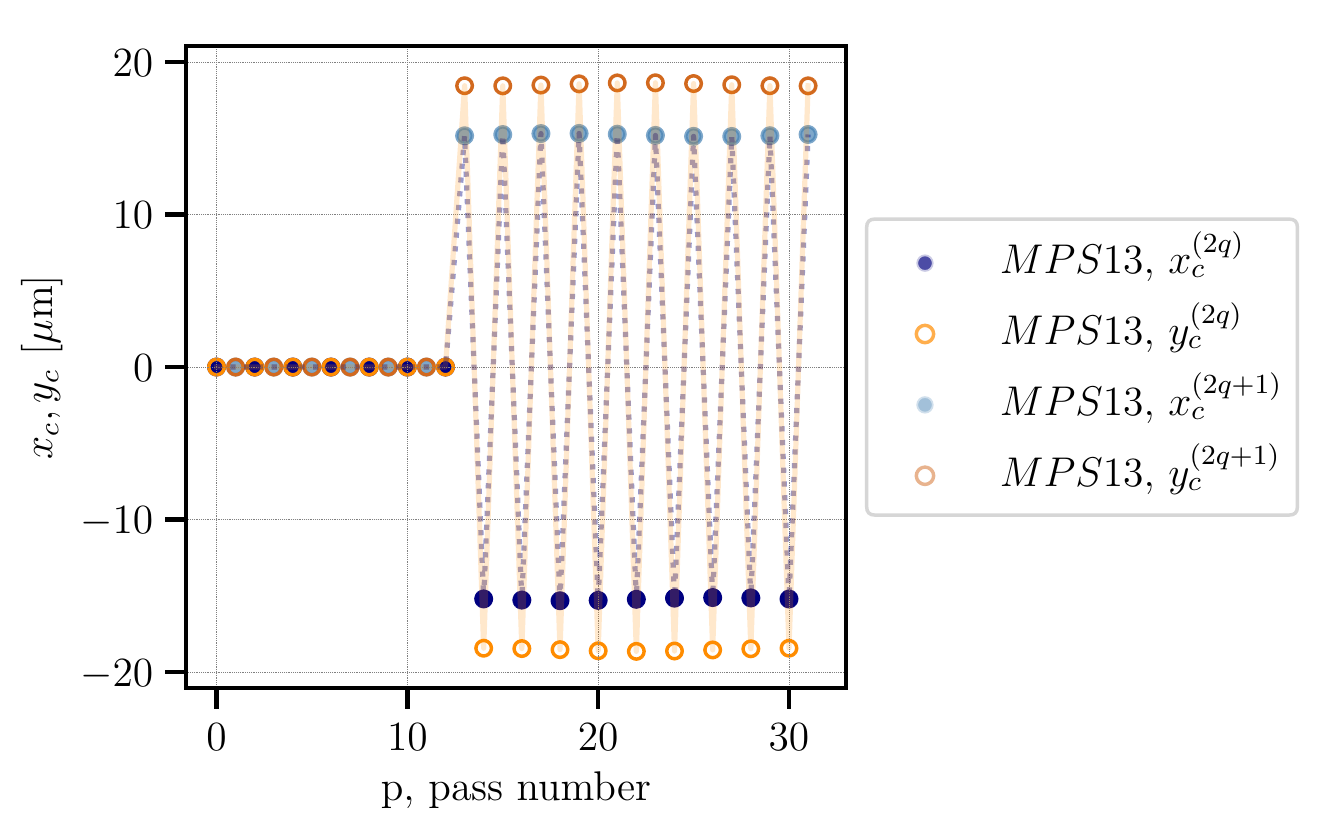}% Here is how to import EPS art
\caption{\label{fig:CIRC-MPS} (Top)-Simulation of the effect of a horizontal parallelism default $d\epsilon_x \sim10\,\mu$m of the MPS$11$ and MPS$13$ in the local referential of the MPS. (Bottom) - The circulating trace resulting from a parallelism default on MPS$13$.}
\end{figure}

A variation of the injection angle $du_x$ ($du_y$) generates two populations of odd and even spot centroids, that are displaced on the $y$ ($x$) axis symmetrically with respect to the position of the spot $p=0$ when no misalignment is made as illustrated in figure  Fig.~\ref{fig:CIRC-M2-ux,uy}. 
% FIGURE MISALIGNMENT INJECTION
\begin{figure}[!htpb]
\includegraphics[width=8cm]{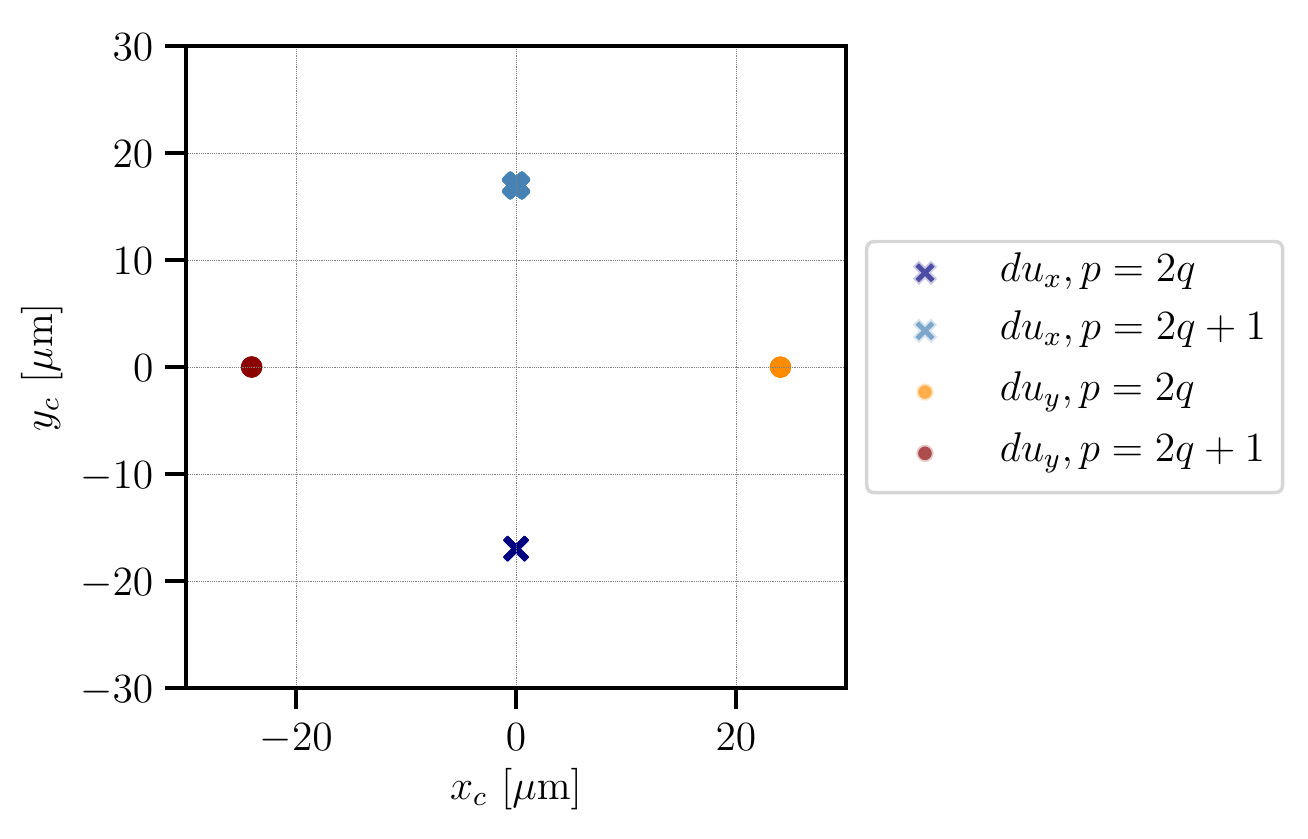}\\% Here is how to import EPS art
\includegraphics[width=8cm]{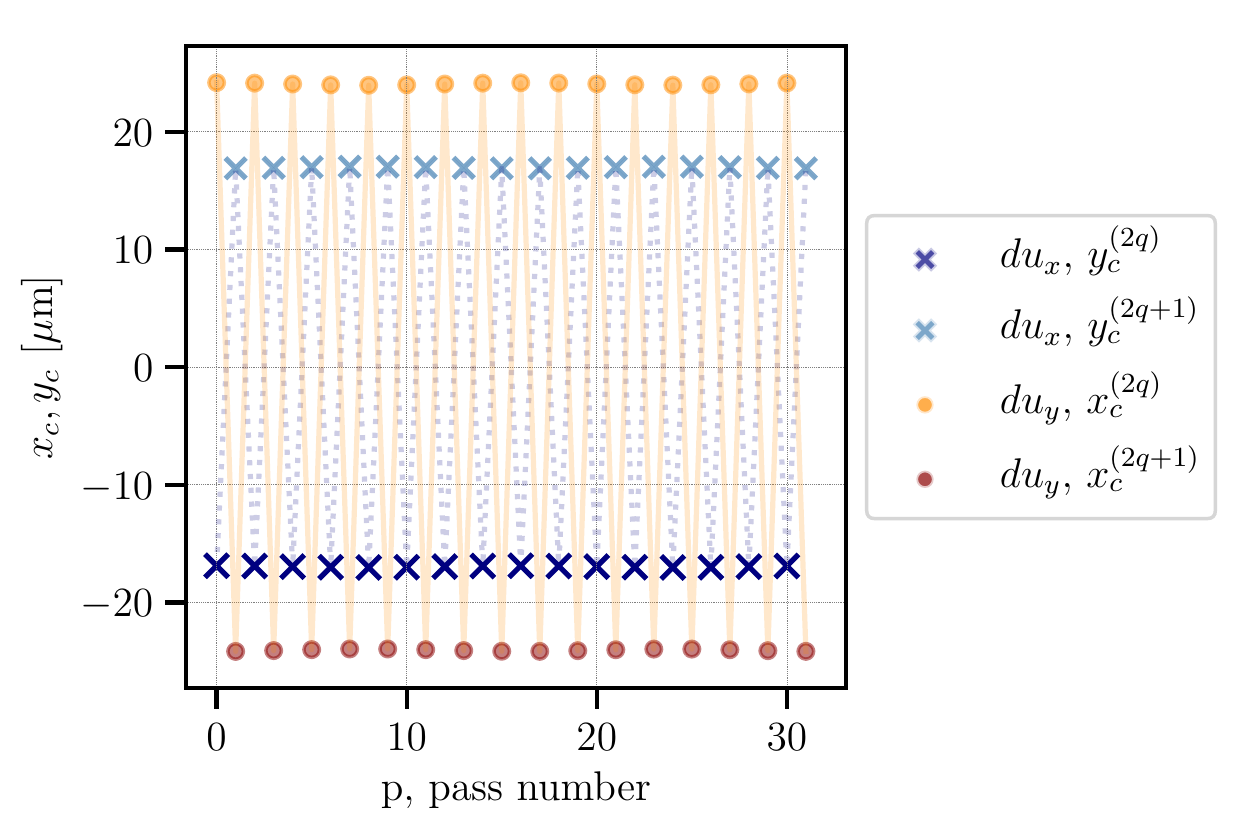}% Here is how to import EPS art
\caption{\label{fig:CIRC-M2-ux,uy} Simulation of the effect of a $du_x = 7.16\,\mu$rad  and  $du_x = -7.16\,\mu$rad pointing angle variation of the injection of the laser beam in the LBC.}
\end{figure}
A MPS misalignment is similar to a combination of an injection misalignments $du_x$ and $du_y$ but creates 3 laser spot populations. The population corresponding to passes $p<k$, before the incriminated MPS$k$ is not modified. Two populations for the subsequent passes are symmetrically created, as if an injection misalignment was present for these passes only. The simulation of the circulation patterns induced by parallelism defects of $\epsilon'_x = 10\,\mu$rad for $MPS11$ and $MPS13$ are displayed in Fig.~\ref{fig:CIRC-MPS} (Top). The simulation of the circulation trace for a parallelism misalignment of  $\epsilon'_x = 10\,\mu$rad of $MPS13$  only is shown in Fig. ~\ref{fig:CIRC-MPS} (Bottom). 

% FIGURE MISALIGNMENT MPS
The typical pattern generated by one of the element of the LBC can help to optimize the alignment process.  As a conclusion,  the laser beam spot centroids misalignment generated pattern and circulation trace,  can be divided in 3 groups:  the odd spots centroid, the even spots centroid and the first pass spot. Using the discrimination in the laser spot centroids the spatial alignment has been completed. 

\subsection{Spatial alignment}

\begin{figure}[!htpb]
\includegraphics[width=8cm]{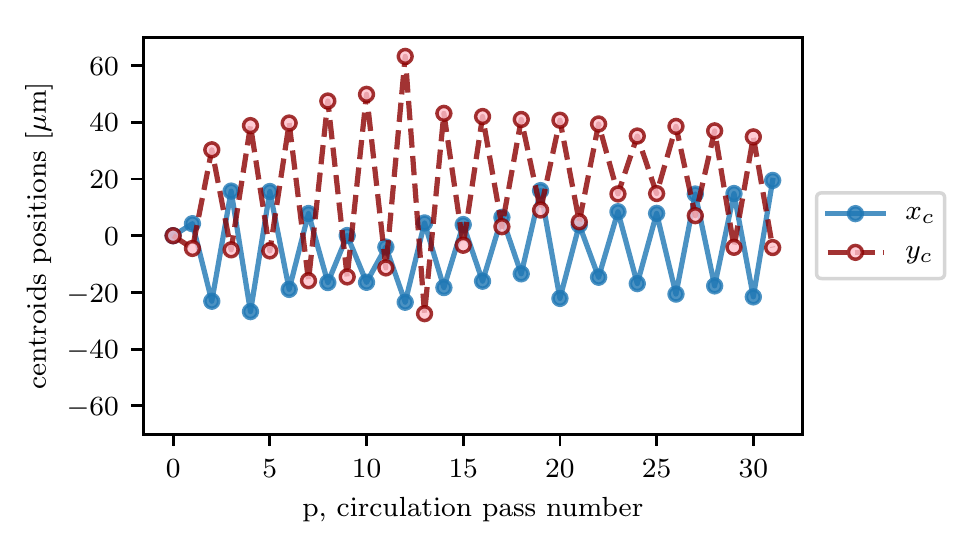}\\% Here is how to import EPS art
\caption{\label{fig:MPS31} Experimental circulation traces obtained by averaging 10 circulation traces after the retuning of the MPS30 and MPS31}
\end{figure}

\begin{figure}[!htpb]
\includegraphics[width = 8cm]{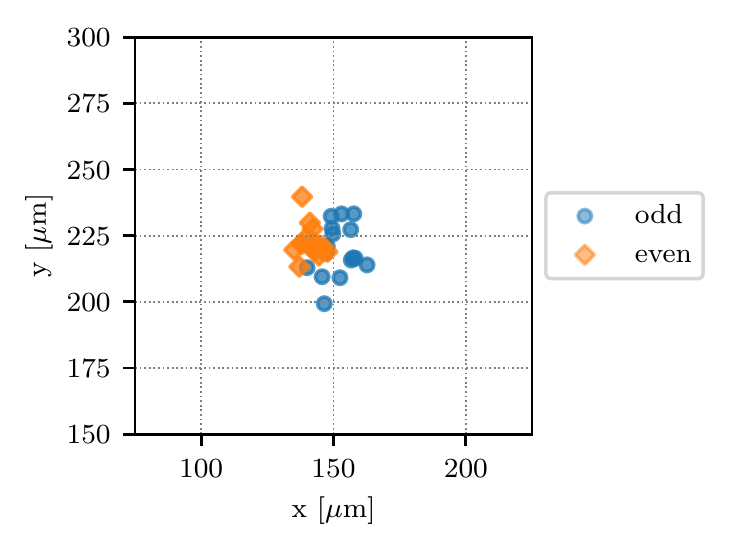}\\% Here is how to import EPS art
\includegraphics[width = 8cm]{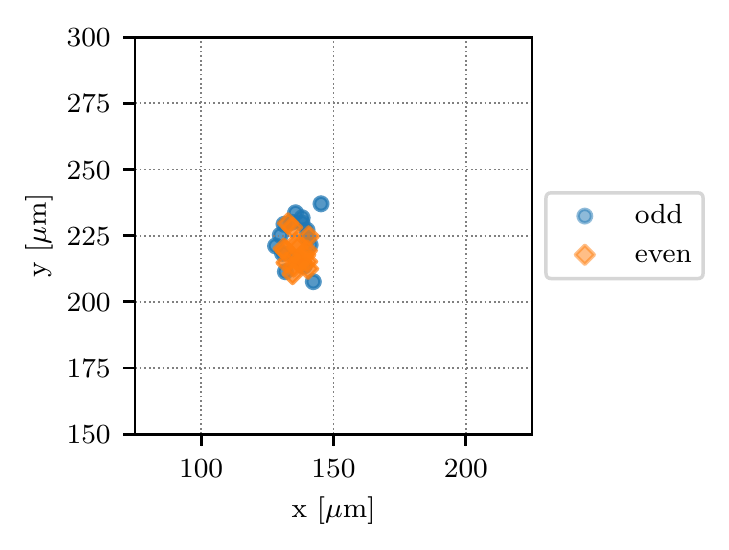}
\caption{\label{fig:injectionMPS1} (top) -  Experimental circulation pattern after the MPS1 steering \textit{in-situ}. (bottom) - Experimental circulation pattern after the fast optimization of the M2 parabola, $RX$, $RY$ and injection of the beam in the LBC}
\end{figure}

\begin{figure}[!htpb] % Here is how to import EPS art
\includegraphics[width=\columnwidth]{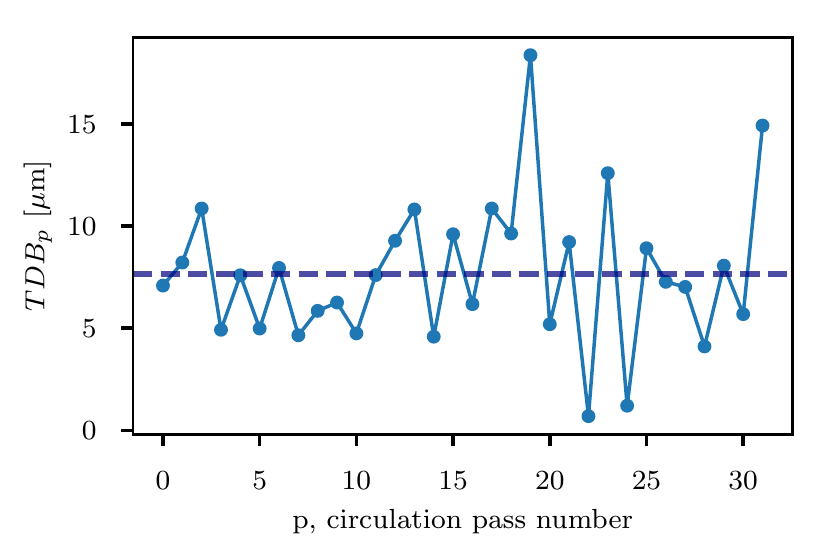}
\caption{\label{fig:TDB} Circulation trace obtained by averaging 10 circulation traces after fast scan optimization on M2.TZ, and laser beam injection. The error in position for each point is $\pm 4\,\mu$m}
\end{figure}

\begin{figure}[!htpb] % Here is how to import EPS art
\includegraphics[width=\columnwidth]{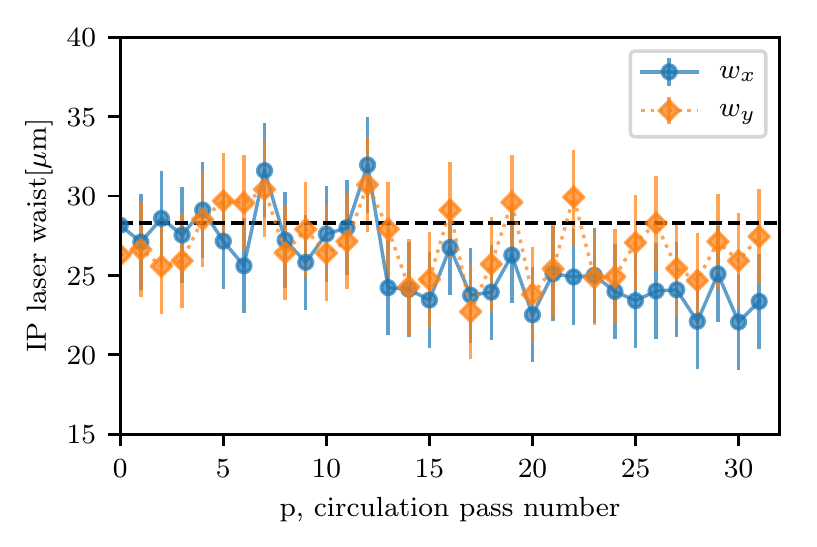}
\caption{\label{fig:beamsizes} The laser-beam waist measured at the IP for each pass after the alignement procedure. The dotted line corresponds to the specified beam size, see Table~\ref{tab:paramIP}.}
\end{figure}

\begin{figure}[!htpb] % Here is how to import EPS art
\includegraphics[width=\columnwidth]{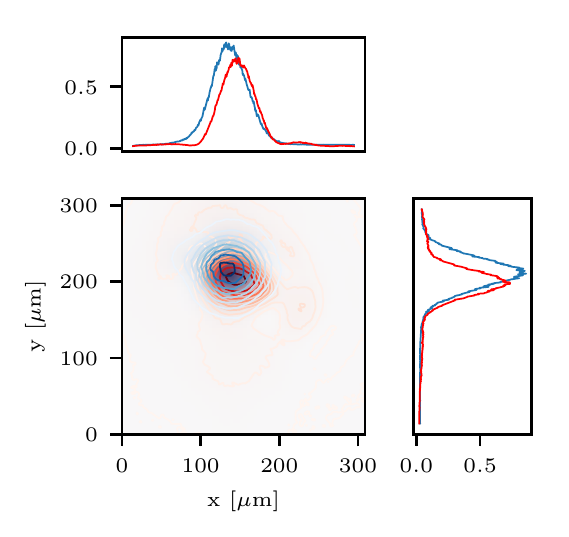}
\caption{\label{fig:pass31} Experimental measurement, the first laser pass (blue) and last pass (red) measured at the IP by the IIOS.}
\end{figure}

Once the MPS installation process is completed the \textit{in-situ} steering of the MPS mirror is done exploiting the behavior of the LBC under misalignments. First the injection is set to get the centroid of pass $p=0$ on the IP reference ($x_0,y_0$) coordinates. The MPS1 is unmounted and controlled on the autocollimator bench and re-installed. Then $RX$ and $RY$ are tilted to super impose the laser spot of pass $p=1$ and on the pass $p=0$. The MPS2 is unmounted, controlled and re-installed. Then the parallelism of all the other MPS are sequentially tuned \textit{in-situ} by superimposing the pass $p$ with the pass $p-2$. At the end of the process the obtained circulation trace is displayed in Fig.~\ref{fig:MPS31}.
Even if the difference is kept below $\sim 10\,\mu$m, between the pass $\#p$ and $\#p-2$ during the in situ alignment process, which corresponds to MPS parallelism alignment error of  $\leq 8\,\mu$rad, we can see that the MPS12 has potentially drifted during the retuning of the next MPS and the pass $\#2$ is not super imposed to the pass $\#0$. This drift is identified to be due to the differential screw shearing.
%\begin{figure}[!htpb]
%\includegraphics[width=8cm]{fig/fig-MPS31-pp-2.png}\\% Here is how to import EPS art
%\caption{\label{fig:MPS31-compp2} Difference between passes $\#p$ and $\#p-2$  $dx_p^{(2)}=x_p-x_{p-2}, dy_p^{(2)} y_p-y_{p-2}$ for $p>1$ .[REDRAW WITHOUT MSP30]}
%\end{figure}
\begin{figure}[!htpb]
\includegraphics[width=8cm]{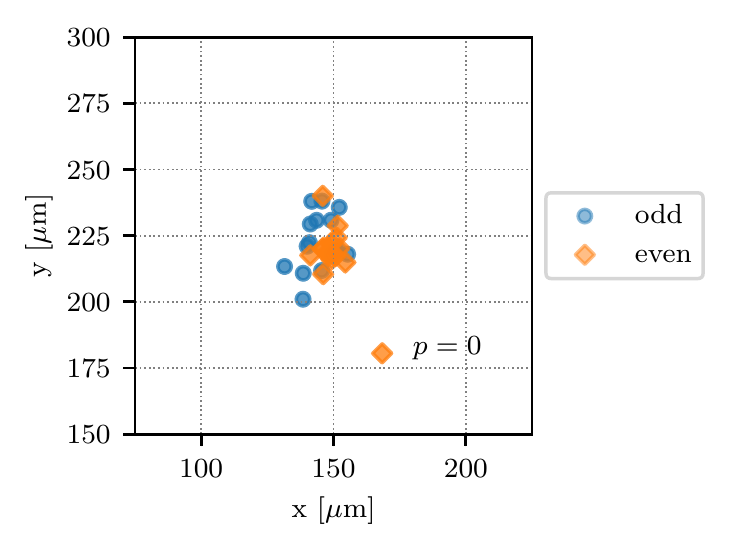}\\% Here is how to import EPS art
\includegraphics[width=8cm]{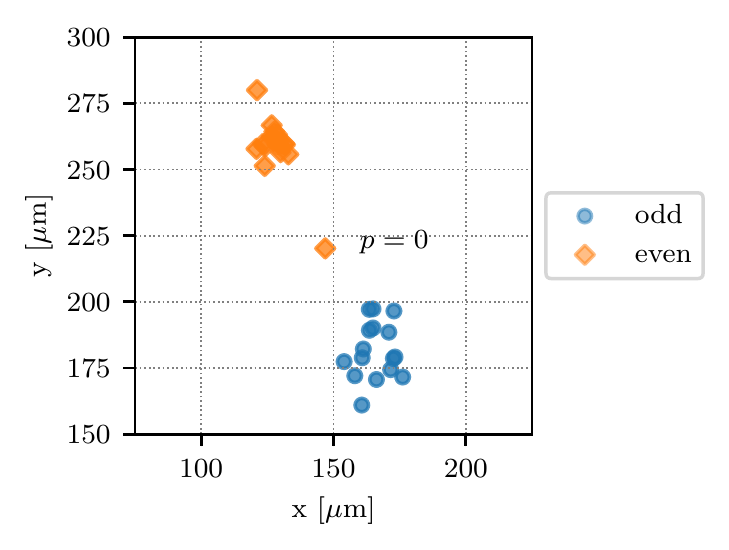}
\caption{\label{fig:MPS31-scatter} (Top)- Experimental circulation pattern obtained after superposition of even and odd groups. (Bottom) - Experimental circulation pattern after injection steering the first pass $p=0$ at the IP position}
\end{figure}
Looking to laser spot centroid at the IP with the IIOS (Fig.~\ref{fig:MPS31-scatter}-top), the two populations (even/odd) of laser spots may be distinguished, the $RX$ and $RY$ rotation are used to overlap the two populations. 
The first pass is then isolated. The injection is steered to get the pass $p=0$ at  $x_0 \approx 148\,\mu$m, $y_0= 223\,\mu$m,  to have $2$ symmetrical populations respect to the pass $p=0$ which correspond to the misalignment error of the MPS1 (Fig.~\ref{fig:MPS31-scatter}-bottom). From the plot of the figure Fig.~\ref{fig:injectionMPS1}  the MPS1 misalignment error is estimated to $\epsilon_x \sim 24\,\mu$rad and $\epsilon_y \sim 38\,\mu$rad. 
MPS1 is retuned {\it in-situ} to overlap the two populations with the first pass $p=0$ as can be seen by comparing the bottom plot of figure Fig.~\ref{fig:MPS31-scatter} and the plot of figure Fig.~\ref{fig:injectionMPS1}. This is the end of the manually driven \textit{in-situ} alignment of the MPS of the LBC. The M2 rotation axes and the injection angles are automatically scanned to minimize the observable introduced in Ref.~\cite{Dupraz:2014aa}, namely the transverse distance to the barycenter (TDB). For the \textit{nth} laser recirculation pass, the distance of its centroid position with respect to the barycenter of the  $32$ laser spot centroid positions is given by:
\begin{equation*}
{\rm TDB}_p=\biggl[\biggl(x_{p}-\frac{1}{32}\sum_{i=1}^{32}x_{i-1}\biggr)^2
					+\biggl(y_{p}-\frac{1}{32}\sum_{i=1}^{32}y_{i-1}\biggr)^2\biggr]^{1/2}.
\end{equation*}
and the average $\overline{TDB}$ of the 32 values of TDB$_n$ yields a performance factor of the alignment. As mentioned in Ref.~\cite{Dupraz:2014aa}, the value of the  $\overline{TDB}$ is related to the relative time average spectral density (TASD$_r$), which is the relevant figure of merit for $\gamma$-ray users \cite{Adriani:2014aa}. 
The values of TDB$_n$ are shown on Fig.~\ref{fig:TDB} along with $\overline{TDB}$. A quick minimization by scanning parameters on M2.$RX,RY, TZ$ and injection beam steering $u_x$,$u_y$, brings the  $\overline{TDB}$ from $14\,\mu$m to $8\,\mu$m at a temperature of $21.1^\circ$C. 
For an optimized alignment the spot sizes are measured for each pass and fitted to gaussian laser spot shape. The average waist are $w_x = 26\pm 3\,\mu$m and $w_y=25\pm3\,\mu$m.  As seen on Fig.~\ref{fig:beamsizes} the beam spot sizes remain almost constant along the laser beam circulation, a slight decrease is however visible. The expected value $w_0 \sim M^{2} \lambda_0 EFL/(\pi w_M) = 1.1 \lambda_0 ROC/(\pi w_M(1+\cos{\Phi})) \sim 25.6\,\mu$m given the beam size $w_M$ on the parabolas, the effective focal length $EFL$  and the $M^{2}$ of the CIRC beam at the exit of the SALT. The dotted value is the waist design value. 
To show the conservation of the good wavefront quality the pass $p=0$ and $p=31$ are shown in the figure Fig.~\ref{fig:pass31}. We can see the diffraction due to the fact that the CIRC beam diameter was slightly to large respect to the max waist design value of the circulating beam at the entrance of the LBC.   

In the final integration of the complete laser system composed by the main high power IP laser and the laser beam transport line up to the interaction point, a fine tuning of the high power laser beam size and high power laser beam divergence is foreseen to accurately tune the laser waist in the LBC. 

%%%
\subsection{Synchronization on the optical reference}

After the LBC alignment is optimized, the laser beam circulation frequency is synchronized on the ELI-NP-GBS reference clock. The ELI-NP-GBS synchronization system has been manufactured by Menlo Systems \cite{Menlo} and consists in a rf reference clock $\nu_{rf} = 2856\,$MHz, a subharmonic clock distribution and a client synchronization. The reference signal is provided by an ultra-low phase noise $\mu$-wave crystal Reference Master Oscillator (RMO) \cite{Piersanti:2018aa}. The absolute jitter is measured below $60\,$fs on the range $10\,$Hz-$10\,$MHz. The RMO provides a stable reference for the Optical Master Oscillator (OMO). The OMO is an Er-doped mode-locked laser oscillator, that encodes the reference timing information in the frequency repetition rate $\nu_{OMO} = \nu_{rf}/46 = 62.086956\,$MHz of short optical pulses at $1560\,$nm. Using such a configuration, the RMO guarantees the long term stability of the OMO, and imprints its low-frequency noise to the whole GBS facility timing.
The OMO pulses are amplified and distributed to all clients. One of the channels is further amplified and converted by a second harmonic generator to a $780\,$nm free space output, used as a seed for the photocathode laser system. Four main clients have been identified in the ELI-NP-GBS machine; two interaction point lasers and two rf extractors (one at 2856 MHz and one at 5712 MHz) that provide the rf reference to the low level rf systems and power units. The clients will be reached by means of four dispersion compensated fiber links with active length stabilization, in order to deliver short sub-$200\,$fs synchronous pulses to the end users. A full description of the synchronization system has been reported in \cite{Piersanti:2016aa}. The frequency of the laser-beam oscillator embedded in the SALT of each IP modules of ELI-NP-GBS is locked on the the reference frequency of the OMO received at the IP location through a non-stabilized fiber link. The synchronization is done thanks to an electro-optical synchronization module developed by Amplitudes Syst\`eme \cite{Casanova:2017aa}. 

During the optical commissioning of the LBC, the whole synchronization system was not available on site. Thus the laser oscillator embedded in the SALT is left free run during operations. Its frequency stability is indeed judged sufficient to allow for the whole synchronization procedure to be done. 

\begin{figure}[!ht] % Here is how to import EPS art
\includegraphics[width=\columnwidth]{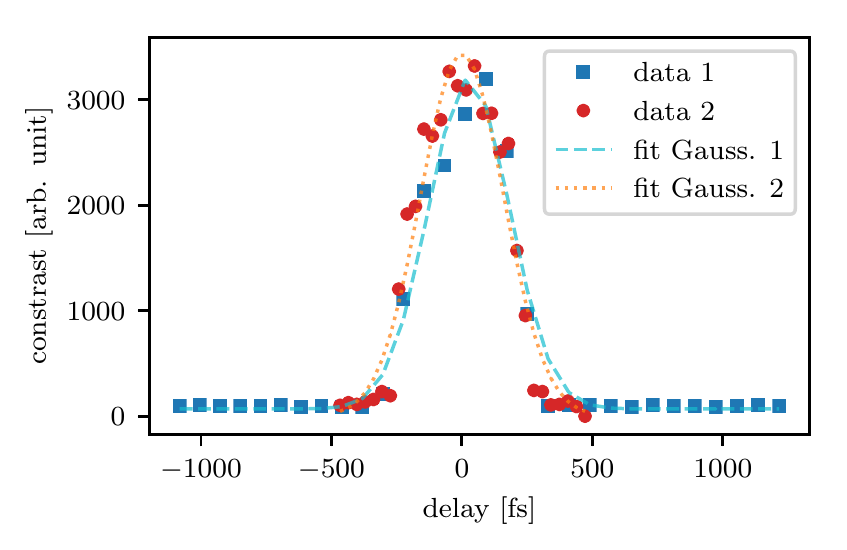}
\caption{\label{fig:syncMPS2} Example of an automated synchronization scan obtained for the MPS3 at to different moments of a day during the commissioning.}
\end{figure}

Using the delay line (DL), see Fig.~\ref{fig:ST}, the first pass $p=0$ of the CIRC beam at the IP and a laser pulse of the REF beam are synchronized at the IP using the first order laser field linear autocorrelation technics described in section \ref{sec:IPmodule} [IIOS]. Then for each pass an automatic scan is realized to acquire the image of the interference pattern as function of the MPS rotation stage position. 
% description algo
For each motor step, several $16$-bit images are acquired with the ICCD detector, each corresponding already to hardware accumulations on the camera. They are then dynamically converted to $8$-bit with respect to the actual measured maximum, blurred with a gaussian with $(5\times5)\,$pixels kernel and dynamically threshold in order to look for lines by means of a \textit{Hough} transform algorithm \cite{Hough:1962aa}. Statistical analysis of the orientation of the lines is used to extract the angle by which the image is then rotated so that the lines become vertical. The lines are summed up and multiplied by a sliding Blackmann-Harris \cite{Harris:1978aa} window of width $150\,$pixels. The result is stored in a two dimensional array that is further discretely Fourier transformed, the magnitude of which is pedestal subtracted. The two main peak magnitudes are averaged to estimate a quantity, proportional to the contrast of the fringes, that is maximized by the automatic scan procedure. An example of the obtained results is given in Fig.~\ref{fig:syncMPS2}. 
%Few parameters had to be tweaked quickly once for all for the system.lue of 0.05*max+0.75*min
%FIG SYNCHRO MPS2/MPS18
The dataset \textit{data1} corresponds to the first synchronization over the full range of the MPS rotation. The second, \textit{data2}, corresponds to a more refined scan taken $3\,$hours after the initial measurement. The synchronization has been observed to be very stable over several days. No corrections on the laser circulation beam path length have been required to keep the accuracy of the synchronisation scan within $\pm 37\,$fs.  Presently the accuracy varies as function of the circulation pass number due to the fact that some passes are less reflected by the TFP, thus degrading the contrast in the ICCD camera. 

\subsection{Stability tests and expected performances}

\begin{figure}[!htpb] % Here is how to import EPS art
\includegraphics[width=\columnwidth]{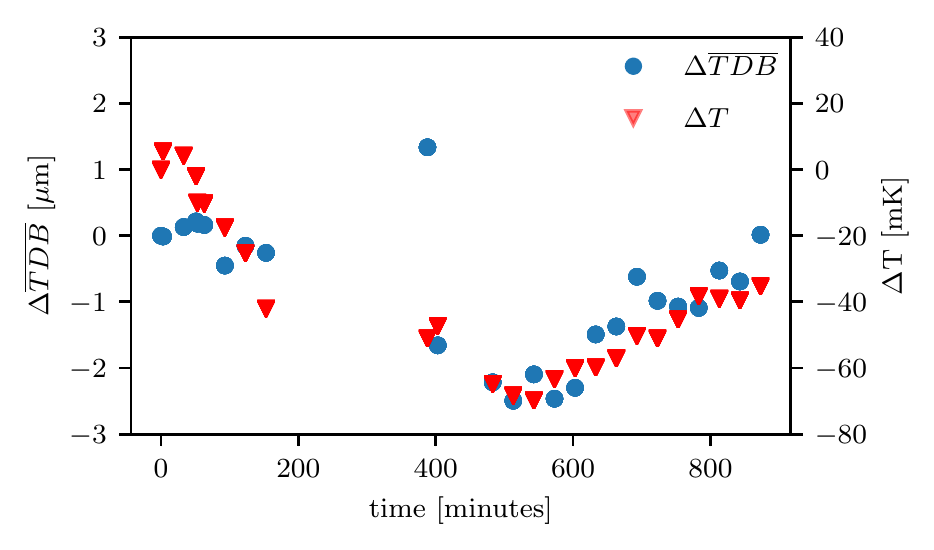}
\caption{\label{fig:LBC-TDB-moni} Variation of the $\overline{TDB}$ and the temperature measured on the MPS bodies under vacuum}
\end{figure}

Once the laser beam circulation is optimized and synchronized, the performances are maintained by locking the injection of the laser ($u_x$,$u_y$) on the injection references (IB.NF2, IB.FF2, see Fig.~\ref{fig:IB}) defined by the alignment optimization and thermalizing the opto-mechanical structure of the LBC to the setting temperature at $\pm 0.1^\circ$C. The system has been pumped down in primary vacuum for the test purpose. 
The stability of the laser beam circulation has been monitored for several days of operation with the low power alignment laser and without active thermalization of the interaction point module. The sensitivity of the system to the temperature has been determined and is shown on Fig.~\ref{fig:LBC-TDB-moni}.
%FIGURE TDB VS TIME 48H VS TEMP VS INJECTION 
%\begin{figure}[!htpb] % Here is how to import EPS art
%\includegraphics[width=8cm]{fig/fig-DTDTDB.pdf}
%\caption{\label{fig:LBC-TDBDT} Variation of $<TDB>$ as a function of the temperature measured on the MPS bodies under vacuum. {\bf FIGURE: replace y et x dans la formule par dTDB et dT}}
%\end{figure}
The stability of the circulation has been monitored regularly over $15$~hours corresponding to a duration larger than a foreseen day of operation. The laser beam injection pointing angle is continuously stabilized on references while the M2 axes have been kept untouched. The variation of the TDB, along with the temperature variation inside vacuum, are presented in Fig.~\ref{fig:LBC-TDB-moni}.

A clear correlation between the $\overline{TDB}$ and the temperature vacuum temperature, which is consistent with that shown on Fig.~\ref{fig:LBC-TDB-moni}.
To ensure the stability of the $\gamma$-ray flux, while considering a perfect electron beam, the temperature stability of the aligned and synchronized LBC must be within $\Delta T < 100\,$mK. A week of operation of the LBC has unveiled that a daily optimization is required in order to ensure that $\overline{TDB} < 15\,\mu$m. 
We experience during the low power commissioning that a starting check of the circulation and a fast optimization (less $15\,$minutes) in the morning was enough to maintain continuous performances for more than $10$h.

\begin{figure}[!ht] % Here is how to import EPS art
\includegraphics[width=\columnwidth]{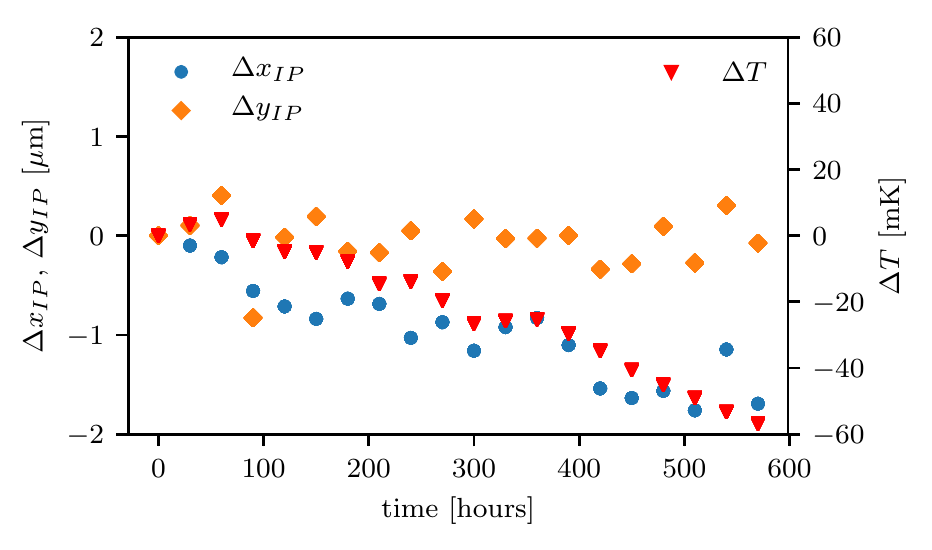}
\caption{\label{fig:IPstab} IP coordinates positions variations $\Delta x_{IP} = x_{IP}(t)-x_{IP}(0)$, $\Delta y_{IP} = y_{IP}(t)-y_{IP}(0)$ during $12\,$hours of laser beam circulation under vacuum. The errors for each IP position point is $\pm1.4\,\mu$m (RMS)}
\end{figure}

The performance obtained during the optical commissioning is considered to be representative of the final operation of the IP module. In fact, for the injection of the high power laser, the only property of the IP module that remains non tested is the laser damage threshold.  However, the laser beam damage threshold for all the optics integrated in the LBC and IB exceed the corresponding requirements, as validated by actual measurements made on witnesses. The M$^2$-quality of the high-power laser-beam has been demonstrated~\cite{Durand:2018aa} independently. It remains to demonstrate the quality of the overlap and synchronization of the electron and laser beams at the IP.  However, tools to achieve this are implemented in the design of the IP module. 
The spatial overlap will be achieved with a micrometer-level precision by means of the OTR light that will be analyzed with the IIOS. The two CBPMs will be used to keep constant the electron beam axis once it is overlapped with the IP defined by the LBC alignment. The $\gamma$-ray flux will be monitored bunch by bunch thanks to a diamond-based fast detector which will allow a precise and instantaneous information about the evolution in time of the system. A feasibility demonstration for the use of such a detector in this context has been made independently \cite{Williams:2016aa}.
Once the electron beam steered at the IP to maximise the $\gamma$-ray flux, it is entirely controlled by the steering magnets of the linac, the beam positions and pointing angles of the high-power laser-beam in the IB. The CBPMs allow to monitor precisely the electron beam axis at the IP. The input (output) positions and pointing laser-beam diagnostics located in the IB (LD) are the only available tools to monitor the high-power laser-beam, the LBC becoming a black-box. The whole opto-mechanical structure, including the quality of the ground floor on which the IP module is set, has to ensure that the spots at the IP are stable within a $\mu$m. It has been demonstrated that this is the case in the hall where the optical commissioning has been conducted. 
The IP coordinates defined as $x_{IP} = \sum_{p=0}^{31} x_p / 32$ and $y_{IP} = \sum_{p=0}^{31} y_p/32$ have been monitored under vacuum and are displayed in Fig.~\ref{fig:IPstab}. 
The IP position remains stable over one work-day of operations. A small drift along the $x$-axis is however noticeable. It is mainly attributed to the temperature variations. Small variations are expected especially due to the thermal deformation of aluminium parts of the parabolic mirror mounts. 

\begin{figure}[!htp] % Here is how to import EPS art
\includegraphics[width=\columnwidth]{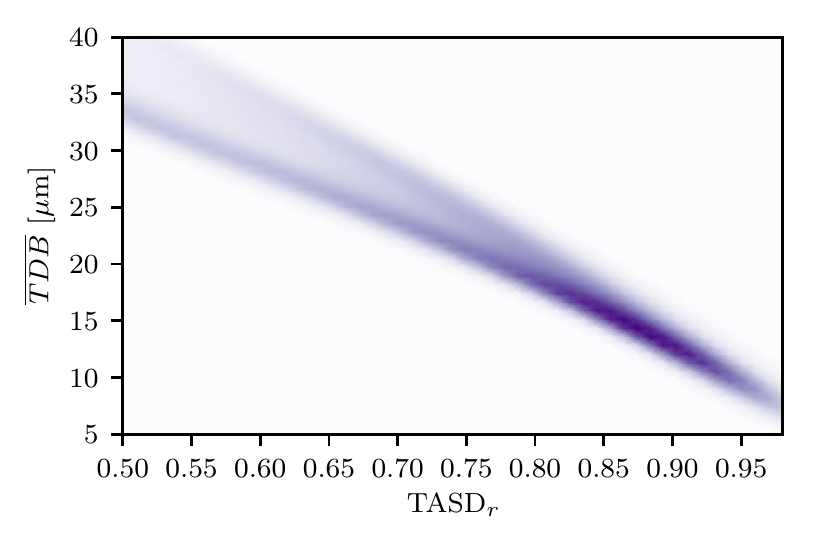}
\caption{\label{fig:TASD-TDB} Computed $\gamma$-ray relative TASD$_r$ as a function of $\overline{TDB}$ at the plane $z=0\,\mu$m}
\end{figure}

The model~\cite{Dupraz:2014aa} is used to simulate $10000$ randomly chosen positions and angles of all the optical elements in the ranges $\pm20\,\mu$m and $\pm20\,\mu$rad, respectively. The gain of the LBC in terms of useful optical power at the IP for the production of $\gamma$-rays, reads:
\begin{equation*}
g(\overline{TDB}) = T_{LBC}\times N\times TASD_r(\overline{TDB}) 
\end{equation*} 
It is roughly estimated as the product of the measured LBC laser beam transmission $T_{LBC} = 0.84\pm0.03$, the number $N$of circulation passes and the relative TASD$_r$. The latter, which represents the loss in spectral density of $\gamma$-rays relative to a perfect theoretical system, is shown on Fig. \ref{fig:TASD-TDB}. 
The best optimization of the alignment experimentally yields a $\overline{TDB}\approx 8\,\mu$m corresponding to a TASD$_r=0.96$ and leading to an effective gain of $g = 25.8\pm0.9$. During days of the commissioning performance tests, and in absence of temperature stabilization of the LBC, the $\overline{TDB}$ is measured to vary between $8\,\mu$m and $15\,\mu$m, which corresponds to $0.85<$TASD$_r<0.93$, thus a relative $5\%$ variation in the gain of the LBC.

\section{\label{sec:Summary} Summary and Outlook} 

The optical commissioning at low power, of the unique opto-mechanical system conceived for the ELI-NP-GBS $\gamma$-ray source, is described. Performances of the procured critical optical elements are given. The setting and integration of the MPS, that are key element of the system, is detailed. The complete alignment procedure has been described and the obtained performances commented. The demonstration of a high stability of the laser beam circulation constitutes in an important achievement towards the installation of the system in the ELI-NP-GBS accelerator hall. It is estimated that the system corresponds to a gain in excess of 25 for the production of $\gamma$-rays by Compton scattering. It demonstrates that the initially foreseen performances of the $\gamma$-beam source 
will be reached for both interaction points. Indeed with the use of the high-power laser-beam of ELI-NP-GBS, delivering $400\,$mJ at $100\,$Hz, an equivalent power of more than $1\,$kW will be used for collisions with the electron beam, delivering a small bandwidth high-flux source. However, it must be underlined that the stable operation of such a highly technological system heavily relies on the good stability of the environnement, in particular the ground stiffness and vibration, the cleanliness of the ambient air, the stability of the temperature and the humidity. Next, a study is on going to improve on the MPS flexor blade mechanical stability before the final integration in the ELINP-GBS tunnel, in order to facilitate the integration process.

\begin{acknowledgments}
We wish to acknowledge EuroGammaS consortium for the support and all the fruitful discussion all along the development of the laser beam circulator. Especially our colleagues from INFN-LNF, INFN-Milano, Universit\`a La Sapienza, Institut Fresnel in Marseille and Amplitude laser company.  The authors would like to acknowledge the efforts of the LAL staff for the continuing support. We also thanks the ALSYOM PUPS group for their help getting the right ambient conditions during the commissioning.
\end{acknowledgments}

%\appendix

\nocite{*}

\bibliographystyle{apsrev}
\bibliography{ndiaye-commissioning-lbc}% Produces the bibliography via BibTeX.

\end{document}